\documentclass[showpacs,floatfix,reprint,pra,superscriptaddress]{revtex4-2}

\usepackage{xfrac}
\usepackage{braket}
\usepackage{graphicx} \graphicspath{{PIC}}
\usepackage{verbatim}
\usepackage{etoolbox}
\usepackage[utf8]{inputenc}
\usepackage{mathtools}
\usepackage{soul,xcolor,colortbl}
\usepackage{hyperref} \hypersetup{colorlinks=true,citecolor=blue,linkcolor=blue,urlcolor=blue}
\usepackage{tabularx}

\usepackage{multirow}
\usepackage{longtable}
\usepackage{cellspace}
\usepackage{resizegather}
\usepackage{dcolumn}
\usepackage[normalem]{ulem}

\usepackage{bm}
\usepackage{array}
\newcolumntype{C}[1]{>{\centering\arraybackslash}p{#1}}\usepackage{soul}
\definecolor{Gray}{gray}{0.85}

\usepackage{color}
\usepackage{morefloats}

\definecolor{Gray}{gray}{0.9}
\definecolor{LightCyan}{rgb}{0.88,1,1}
\definecolor{green}{rgb}{0.5451,0.2706,0.0745}

\def\DFT{{\small  DFT}}
\def\CCE{{\small  CCE}}
\def\LDA{{\small  LDA}}
\def\GGA{{\small  GGA}}
\def\PBE{{\small  PBE}}
\def\SCAN{{\small  SCAN}}
\def\FERE{{\small  FERE}}
\def\MAE{{\small  MAE}}
\def\MAEs{{\small  MAE}s}

\def\ZPC{{\small  ZPC}}
\def\TC{{\small  TC}}
\def\TVC{{\small  TVC}}
\def\NISTJANAF{{\small  NIST-JANAF}}
\def\ICSD{{\small  ICSD}}
\def\AFLOWSYM{{\small  AFLOW-SYM}}

\def\AGL{{\small AGL}}

\def\AFLOW{{\small AFLOW}}
\def\VASP{{\small VASP}}

\def\sDFT{{\substack{\scalebox{0.6}{DFT}}}}

\def\svib{{\substack{\scalebox{0.6}{vib}}}}

\def\sformula{{\substack{\scalebox{0.6}{$A_{x_1}\mathrm{\dots }Y_{x_n}$}}}}

\makeatletter \renewcommand\frontmatter@abstractwidth{\dimexpr\textwidth\relax} \makeatother

\def\citePROTOS{\cite{curtarolo:art121,curtarolo:art145,curtarolo:art173}}
\def\tablecolumnspacing{2.3}

\begin{document}

\title{AFLOW-CCE for the thermodynamics of ionic materials}

\author{Rico Friedrich}
\email[]{rico.friedrich@tu-dresden.de}
\affiliation{Theoretical Chemistry, Technische Universität Dresden, 01062 Dresden, Germany}
\affiliation{Institute of Ion Beam Physics and Materials Research, Helmholtz-Zentrum Dresden-Rossendorf, 01328 Dresden, Germany}
\affiliation{Center for Autonomous Materials Design, Duke University, Durham, NC 27708, USA}
\author{Stefano Curtarolo}
\email[]{stefano@duke.edu}
\affiliation{Center for Autonomous Materials Design, Duke University, Durham, NC 27708, USA}
\affiliation{Materials Science, Electrical Engineering, and Physics, Duke University, Durham, NC 27708, USA}

\date{\today}

\begin{abstract}
\noindent
Accurate thermodynamic stability predictions enable data-driven computational materials design.
Standard density
functional theory (\DFT) approximations have limited accuracy with average errors of a few hundred meV/atom for ionic materials such as oxides and nitrides.
Thus, insightful correction schemes as given by the coordination corrected enthalpies (\CCE) method, based on an intuitive parameterization of \DFT\ errors with respect to coordination numbers and cation oxidation states present a simple, yet accurate solution to enable materials stability assessments.
Here, we illustrate the computational capabilities of our \AFLOW-\CCE\ software by utilizing our previous results for oxides and introducing new results for nitrides.
The implementation reduces the deviations between theory and experiment to the order of the room temperature thermal energy scale, \emph{i.e.} $\sim$25 meV/atom.
The automated corrections for both materials classes are freely available within the \AFLOW\ ecosystem via the \AFLOW-\CCE\ module, requiring only structural inputs.

\vspace{0.2cm}

\end{abstract}

\maketitle

\noindent
In data-driven computational materials science, density functional theory (\DFT) has become the standard method to characterize systems and an indispensable tool enabling insightful predictions.
Over the last decades, John Perdew and his group have made \DFT\ accurate enough for materials design while setting the standards in the field according to the functional developments related to \LDA~\cite{LDA}, \PBE~\cite{PBE}, and \SCAN~\cite{Perdew_SCAN_PRL_2015}.
These approximations are used for the data generation of virtually all big materials databases with millions of entries
\cite{Oses_CMS_2023,Esters_CMS_2023,curtarolo:art75,materialsproject.org,oqmd.org,Kirklin_NPJCM_2015,nomadMRS,ase,cmr_repository,Pizzi_AiiDA_2016}, and thus closely match the quest posed by Dirac for ``approximate practical methods'' to solve the quantum many-body solid state problem~\cite{Dirac1929}.

Thermodynamics is one of the most crucial aspects for materials design, since synthesizability can be ensured for thermodynamically stable systems.
As such, the need for an accurate computational treatment of thermodynamic stability is pressing for many issues related to \emph{e.g.} stability of competing phases~\cite{curtarolo:art176}, high-entropy materials~\cite{SarkerHarrington_HEC_NCOMMS_2018_etal,Oses_NatureReview_2020}, solution enthalpies in liquid metals~\cite{Gil_PCCP_2022}, and novel (2D) nanostructures when evaluating relative energies and with respect to the bulk parent compound~\cite{Puthirath_Balan_NNANO_2018,Friedrich_NanoLett_2022,Barnowsky_AdvElMats_2023,Balan_MatTod_2022,Kaur_AdvMat_2022}.
The formation enthalpy rigorously quantifies the thermodynamic stability, as the enthalpy difference between the material and its elemental references.
Its accurate calculation is a basic challenge.
Standard (semi-)local and even currently available more advanced \emph{ab initio} methods yield errors of several hundred meV/atom for ionic compounds such as oxides and nitrides~\cite{Wang_Ceder_GGAU_PRB_2006,Lany_FERE_2008,Jain_GGAU_PRB_2011,Stevanovic_FERE_2012,Yan_formation_PRB_2013,Jauho_PRB_2015,Zhang_NPJCM_2018,Isaacs_PRM_2018,Friedrich_CCE_2019} when comparing to validated standard collections for thermochemical data~\cite{Kubaschewski_MTC_1993,Chase_NIST_JANAF_thermochem_tables_1998,Barin_1995,Wagman_NBS_thermodyn_tables_1982}.
The reason is that accurate total energies for all systems involved --- compound and elemental references --- are still not accessible with the standard functionals, and errors when calculating the enthalpy difference between chemically dissimilar systems do not cancel.
The explicit treatment of self-interaction errors might be an avenue to further significant advances from \emph{ab initio}~\cite{Pederson_JCP_2014,Yang_PRA_2017,Kao_JCP_2017,Schwalbe_JCC_Fermi-Lowdin_2018}, but likely at an elevated computational cost infeasible for high-throughput materials design.
Even efforts based on Quantum Monte Carlo calculations only partially remedy the discrepancies between theory and experiment~\cite{Pozzo_PRB_2008,Mao_QMC_2011,Isaacs_PRB_2022}.

A solution for this dilemma can be provided by physically motivated empirical enthalpy corrections.
These parameterize \mbox{(semi-)local} \DFT\ errors with respect to measured values, are hence feasible to enable materials design, and can include uncertainty quantification~\cite{Wang_SciRep_2021}.
It turns out that correction methods, mixing schemes, and machine learning approaches based on only the composition of a material can already lead to significant improvements in accuracy reducing the \underline{m}ean \underline{a}bsolute \underline{e}rrors (\MAEs) down to $\sim$50~meV/atom~\cite{Wang_Ceder_GGAU_PRB_2006,Lany_FERE_2008,Jain_GGAU_PRB_2011,Stevanovic_FERE_2012,Wolverton_DFTUenthalpies_prb_2014,Artrith_PRM_2022,Kingsbury_NPJCM_2022,Gong_JACSAu_2022}.
In addition to the still limited accuracy, such an approach can, however, be problematic for the phase diagrams of certain systems due to only tilting the Gibbs energy landscape~\cite{Friedrich_CCE_2019}.
These schemes do also not allow for a correction of the \emph{relative} stability of polymorphs whose energetic ordering is known to be incorrectly predicted by \DFT\ in several cases~\cite{Zhang_NPJCM_2018}.

To rectify these issues, we have recently introduced the method of \underline{c}ordination \underline{c}orrected \underline{e}nthalpies (\CCE), based on the bonding topology in a material~\cite{Friedrich_CCE_2019}.
\CCE\ avoids thermodynamic paradoxes by construction, reduces the \MAE\ down to $\sim$25~meV/atom, and can correct the relative stability of polymorphs if they differ in the number of nearest neighbor bonds~\cite{Friedrich_CCE_2019, Friedrich_PRM_2021}.
So far, the method has been parameterized for oxides and a few halides calling for an extension to other anion classes such as nitrides.
Most importantly, effective implementations are needed freely available to and easy to use by the computational materials science community.
Here, we provide such a tool by our automated corrections within the \AFLOW-\CCE\ module.

\begin{figure*}[ht!]
	\centering
	\includegraphics[width=0.7\textwidth]{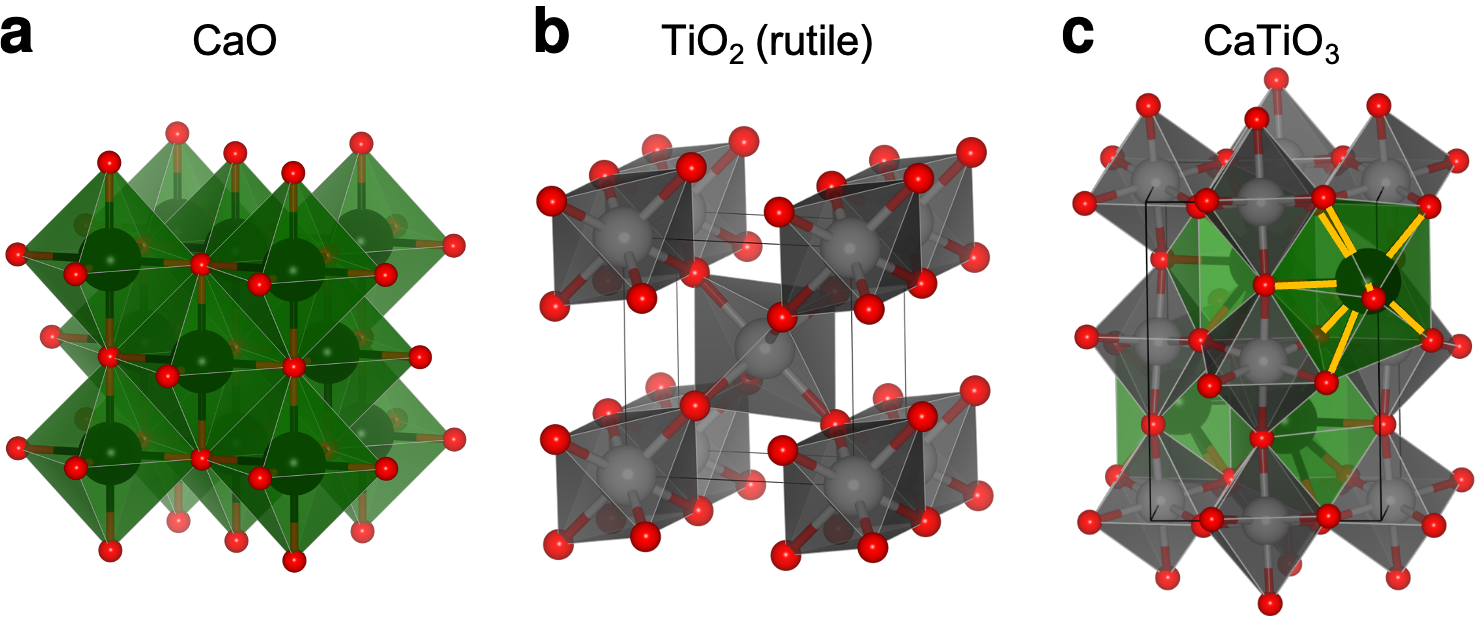}
	\caption{\small \textbf{Corrections per bond.}
	Structural models of (\textbf{a}) CaO, (\textbf{b}) rutile TiO$_2$, and (\textbf{c}) perovskite CaTiO$_3$.
	While the Ti coordination number is six in both rutile and perovskite, Ca exhibits a coordination change from sixfold to eightfold when going from CaO to CaTiO$_3$ (Ca–O bonds marked in yellow).
	The number of bonds is thus a crucial parameter for the thermodynamic stability of a material.
    }
	\label{fig0}
\end{figure*}

In this article, we {\bf i.} present some theoretical basics, a motivation of \CCE, and its formalism; and {\bf ii.} demonstrate the computational capabilities of our implementation using our previous results for oxides and new results for nitrides, including a discussion of the error bars of plain \DFT\ for different standard functionals, as well as a demonstration of the performance of the \AFLOW-\CCE\ module for a test set of ternary nitrides.

\section*{Theory} \label{theory}
\noindent
{\bf Formation enthalpies}. From \DFT, one directly computes an estimate of the formation enthalpy $\Delta_{\mathrm{f}} E^{0,\mathrm{{\sDFT}}}$ of a compound $A_{x_1}B_{x_2}\dots Y_{x_n}$ at zero temperature and pressure without zero-point vibrations:

\begin{equation*}\label{dft_form_energy}
\Delta_{\mathrm{f}} E^{0,\mathrm{{\sDFT}}}_{\sformula}=U^{0,\mathrm{{\sDFT}}}_{\sformula}- \left[\sum_{i=1}^{n-1} x_i U^{0,\mathrm{{\sDFT}}}_{i} + \frac{x_{n}}{2}U^{0,\mathrm{{\sDFT}}}_{Y_{2}}\right],
\end{equation*}

\noindent
where $U^{0,\mathrm{{\sDFT}}}_{\sformula}$, $U^{0,\mathrm{{\sDFT}}}_{i}$, and $U^{0,\mathrm{{\sDFT}}}_{Y_{2}}$ are the total internal energies of the compound, the $i$-element reference phase, and molecular $Y_2$, respectively; and $x_1, ... , x_n$ are stoichiometries.
In this manuscript, $Y$ stands for O or N and this notation can be extended to other anion species.

On the other hand, the tabulated measured standard (``$\circ$'') formation enthalpy at room temperature ($T_{\mathrm{r}}=$298.15~K) corresponds to:

\begin{equation*}\label{standard_form_enth_def}
\Delta_{\mathrm{f}} H^{\circ,T_{\mathrm{r}},\mathrm{exp}}_{\sformula}=H^{\circ,T_{\mathrm{r}}}_{\sformula}- \left[\sum_{i=1}^{n-1} x_i H^{\circ,T_{\mathrm{r}}}_{i} + \frac{x_{n}}{2}H^{\circ,T_{\mathrm{r}}}_{Y_{2}}\right],
\end{equation*}

\noindent
where $H^{\circ,T_{\mathrm{r}}}_{\sformula}$, $H^{\circ,T_{\mathrm{r}}}_{i}$, and $H^{\circ,T_{\mathrm{r}}}_{Y_{2}}$ are the standard enthalpies of the compound, the $i$-element reference phase, and $Y_2$ at $T_{\mathrm{r}}$.

As we have shown in Ref.~\onlinecite{Friedrich_CCE_2019}, the measured value can be written in a very good approximation (accuracy better than 1~meV/atom) as the sum of the internal energy contribution and a small vibrational term $\Delta_{\mathrm{f}} H^{\mathrm{\svib}}_{\sformula}$ of the order of 10-20~meV/atom due to zero-point and thermal effects:

\begin{equation}\label{standard_form_enth_approx2}
	\Delta_{\mathrm{f}} H^{\circ,T_{\mathrm{r}},\mathrm{exp}}_{\sformula} \approx \Delta_{\mathrm{f}} E^{0}_{\sformula} + \Delta_{\mathrm{f}} H^{\mathrm{\svib}}_{\sformula} \approx \Delta_{\mathrm{f}} H^{\circ,T_{\mathrm{r}},\mathrm{cal}}_{\sformula}.
\end{equation}

\noindent
Both contributions in Eqn.~(\ref{standard_form_enth_approx2}) can be estimated computationally to yield $\Delta_{\mathrm{f}} H^{\circ,T_{\mathrm{r}},\mathrm{cal}}_{\sformula}$, the calculated standard formation enthalpy at $T_{\mathrm{r}}$~\cite{Friedrich_CCE_2019}.

The vibrational term can be computed reliably within a quasi-harmonic Debye model as implemented within the AFLOW \underline{A}utomatic \underline{G}IBBS \underline{L}ibrary (\AGL)~\cite{Blanco_jmolstrthch_1996,BlancoGIBBS2004,curtarolo:art96,curtarolo:art115,Poirier_Earth_Interior_2000,curtarolo:art125} with an average error as small as $\sim$5~meV/atom~\cite{Friedrich_CCE_2019}.
The zero-point and thermal effects of O$_2$/N$_2$ are accounted for based on reliable tabulated (spectroscopic) data~\cite{Chase_NIST_JANAF_thermochem_tables_1998} as well as assuming perfect gas behavior.

The large errors between calculated and measured formation enthalpies are due to estimating the internal energy contribution in
Eqn.~(\ref{standard_form_enth_approx2}) from \DFT.

\noindent
{\bf Motivation of corrections per bond.} To obtain accurate enthalpies, an insightful correction scheme is required.
The idea behind \CCE\ is illustrated in Fig.~\ref{fig0}.
When forming perovskite CaTiO$_3$ out of CaO and rutile TiO$_2$, the number of nearest neighbor cation-anion bonds, \emph{i.e.} the coordination number of Ca, changes from six to eight.
The number of bonds in a material is crucial for its thermodynamic stability and it is commonly discussed how well a specific approximation to \DFT\ captures a certain type of bonding~\cite{Sun_NChem_2016}.
It thus seems plausible to assign an error per bond to the computed enthalpies making the coordination number an appropriate descriptor to parameterize \DFT\ errors.
For optimal transferability from fit to target compounds, the correction also needs to be specific to the oxidation number of the cation as a different self-interaction error is expected for different charge states of transition metal centers such as Fe$^{2+}$ \emph{vs.} Fe$^{3+}$.

\noindent
{\bf The \CCE\ formalism.}
Taking binary compounds $A_{x_1}Y_{x_2}$ as the fit set, the \CCE\ corrections $\delta H^{T,A^{+\alpha}}_{A-Y}$ per cation-anion $A-Y$ bond and cation oxidation state $+\alpha$ are obtained from the difference between \DFT\ formation enthalpies and experimental standard formation enthalpies at temperature $T$~\cite{Friedrich_CCE_2019}:

\begin{equation*}\label{fit_CCE}
\Delta_{\mathrm{f}} E^{0,\mathrm{DFT}}_{A_{x_1}Y_{x_2}}-\Delta_{\mathrm{f}} H^{\circ,T,\mathrm{exp}}_{A_{x_1}Y_{x_2}} =x_1N_{A-Y}\delta H^{T,A^{+\alpha}}_{A-Y},
\end{equation*}

\noindent
where $N_{A-Y}$ is the number of nearest neighbor $A-Y$ bonds and $x_{i}$ are stoichiometries for the $i$-species.
$T$ can be 298.15 or 0~K, \emph{i.e.} temperature effects are included in the corrections as 0~K estimates of experimental $\Delta_{\mathrm{f}}H$ values can be obtained by taking into account thermal effects only for the fit compounds according to Ref.~\cite{Friedrich_PRM_2021}.
As detailed in Refs.~\cite{Friedrich_CCE_2019,Friedrich_PRM_2021}, the corrections are fit to the formation enthalpies directly calculated from \DFT, without taking into account thermal/vibrational contributions since these terms are to a very good approximation (average accuracy $\sim$1~meV/atom) implicitly included in the corrections.

The corrections can then be applied without additional computational cost compared to plain \DFT\ to any multinary compound $A_{x_1}B_{x_2}\dots Y_{x_n}$ to obtain the \CCE\  formation enthalpy $\Delta_{\mathrm{f}} H^{\circ,T,\mathrm{CCE}}_{A_{x_1}B_{x_2}\mathrm{\dots}Y_{x_n}}$:

\begin{equation*}\label{apply_CCE}
\Delta_{\mathrm{f}} H^{\circ,T,\mathrm{CCE}}_{A_{x_1}B_{x_2}\mathrm{\dots}Y_{x_n}} =\Delta_{\mathrm{f}} E^{0,\mathrm{DFT}}_{A_{x_1}B_{x_2}\mathrm{\dots}Y_{x_n}} - \sum_{i=1}^{n-1} x_iN_{i-Y}\delta H^{T,i^{+\alpha}}_{i-Y},
\end{equation*}

\noindent
where $N_{i-Y}$ is the number of nearest neighbor bonds between the cation $i$ and anion $Y$-species.
The method has been implemented for automated enthalpy corrections requiring only an input structure as the \AFLOW-\CCE\ module available to the scientific community~\cite{Friedrich_PRM_2021}.
The implementation is also applicable to enthalpies computed from \DFT$+U$ if the same settings as for the \AFLOW-{\small ICSD} database are used~\cite{curtarolo:art104, Friedrich_PRM_2021}.

The \emph{ab-initio} calculations of this study for the exchange-correlation functionals \LDA~\cite{DFT,LDA,von_Barth_JPCSS_LSDA_1972}, \PBE~\cite{PBE}, and \SCAN~\cite{Perdew_SCAN_PRL_2015} are performed with \AFLOW~\cite{curtarolo:art53,curtarolo:art57,curtarolo:art63,aflowPAPER,curtarolo:art110,aflowPI,Esters_CMS_2023,Oses_CMS_2023} and the {V}ienna \emph{{A}b-initio} {S}i\-mu\-lation {P}ackage (\VASP)~\cite{vasp} with settings according to Refs.~\onlinecite{Friedrich_CCE_2019,curtarolo:art104}.

\section*{Results and Discussion} \label{results_discussion}

\begin{figure*}[ht!]
	\centering
	\includegraphics[width=\textwidth]{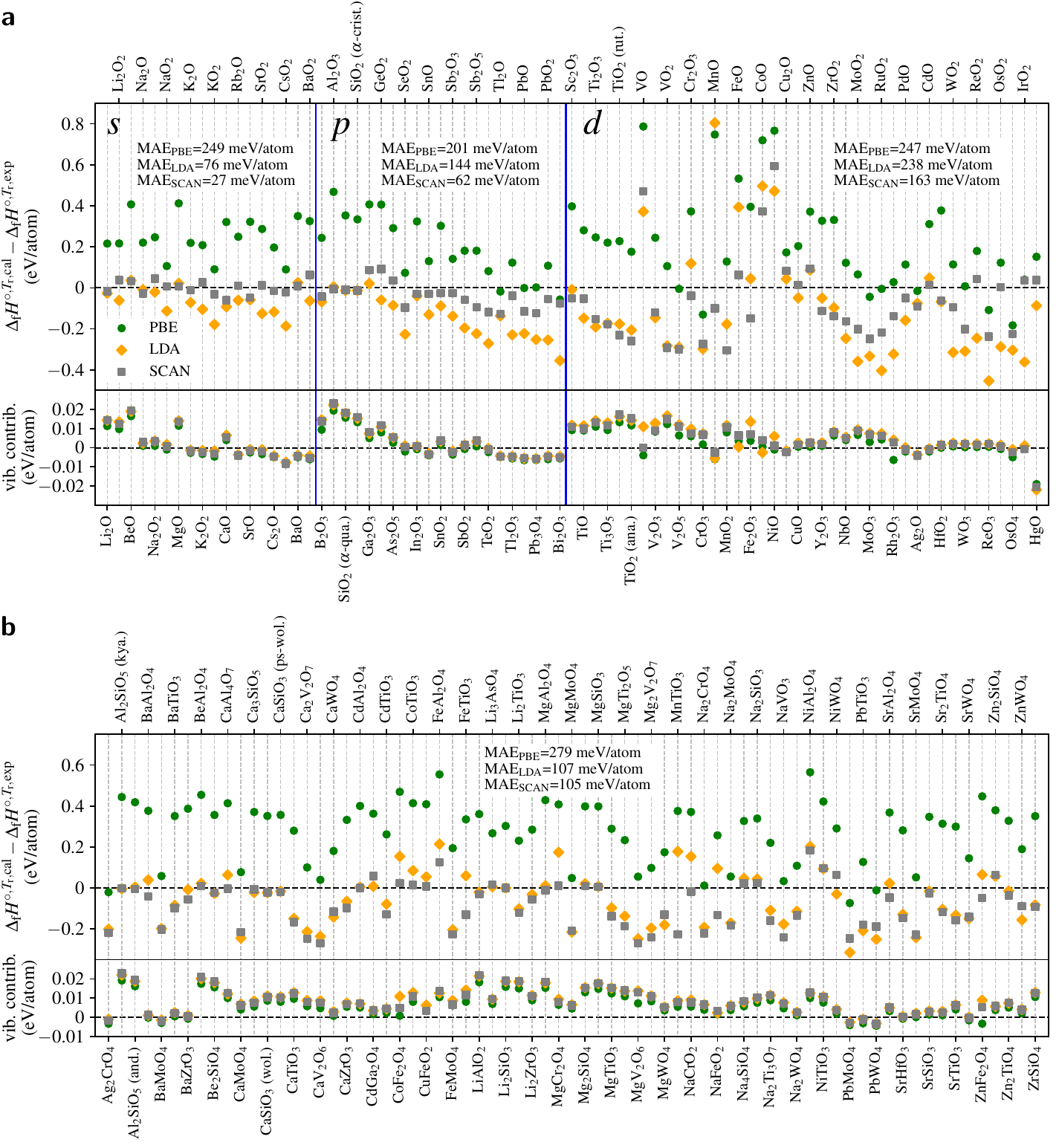}
	\caption{\small \textbf{Formation enthalpies and vibrational contribution for binary oxides.}
	(\textbf{a}) Differences between calculated and experimental room temperature formation enthalpies of 79 binary oxides (upper panel) and vibrational (zero-point + thermal) contribution to the calculated formation enthalpy (lower panel) for three standard \DFT\ functionals.
    Vertical blue lines separate blocks with different $l$-character of the cations.
    (\textbf{b}) Differences between calculated and experimental room temperature formation enthalpies of 71 ternary oxides (upper panel) and vibrational contribution (lower panel).
    Adapted from
    Ref.~\onlinecite{Friedrich_CCE_2019}.
    }
	\label{fig1}
\end{figure*}

\noindent
{\bf Implementation: corrections for oxides.}
We first demonstrate the capabilities of our \AFLOW-\CCE\ implementation on our previous data for oxides included in the Supporting Information of Ref.~\onlinecite{Friedrich_CCE_2019}.
While absolute formation enthalpies are on the order of several eV/atom, Fig.~\ref{fig1} depicts the difference between computed (\DFT+\AGL) and measured room temperature formation enthalpies for 79 binary (\textbf{a}) and 71 ternary (\textbf{b}) oxides also including a visualization of the vibrational contribution in the lower panel in both cases.
The errors for all functionals are substantial with \SCAN\ delivering the most accurate results still showing overall \MAEs\ of at least about 100~meV/atom.
The \MAEs\ over the whole binary set are 235, 176, and 105~meV/atom for \PBE, \LDA, and \SCAN, respectively.

In general, \PBE\ underestimates the (absolute) formation enthalpies meaning that they are less negative than the experimental values while \LDA\ and \SCAN\ mostly overestimate them.
For the binaries, the $l$-character of the cation has, however, a strong influence.
While \SCAN\ yields constantly very accurate enthalpies with a \MAE\ as small as 27~meV/atom for $s$-element oxides, all functionals show a systematic decrease of the computed values for heavier $p$-oxides which can be related to increasing covalency~\cite{Friedrich_CCE_2019}.
For the $d$-oxides, the systematic trends of the functionals are diminished and large individual errors ranging up to $\sim$800~meV/atom are observed increasing also the \MAEs\ to over 150~meV/atom.
This drastic behavior for the rock salt structure $d$-oxides VO, MnO, FeO, CoO, and NiO is well known~\cite{Stevanovic_FERE_2012,Friedrich_CCE_2019}.
For the ternaries, mostly including transition metal elements, the large errors are confirmed with \MAEs\ of at least 100~meV/atom for each functional.

The vibrational contribution is very small in all cases reaching at max about 20~meV/atom for Al$_2$O$_3$ and kyanite Al$_2$SiO$_5$ and as such are about two orders of magnitude smaller than absolute formation enthalpies.

\begin{figure*}[ht!]
	\centering
	\includegraphics[width=\textwidth]{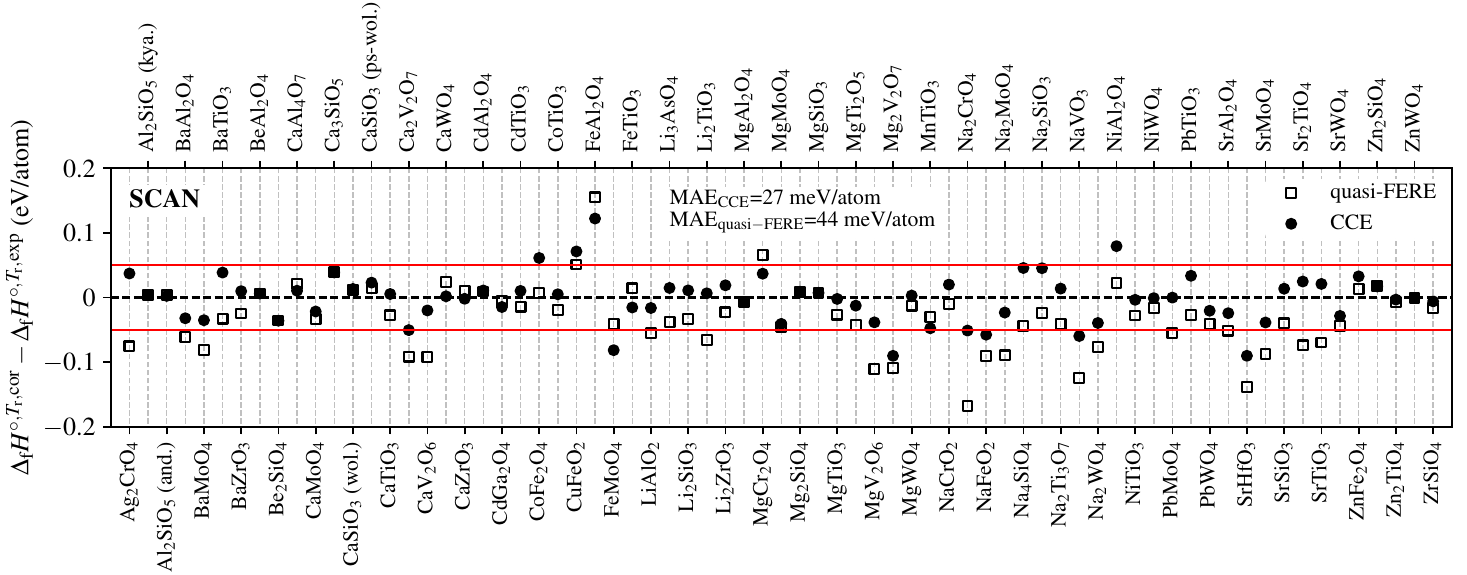}
	\caption{\small \textbf{Corrected \SCAN\ enthalpies for ternary oxides.}
	Differences between corrected and experimental room temperature formation enthalpies of 71 ternary oxides for \SCAN.
	In addition to the \CCE\ results, a comparison to an implementation of the \FERE\ method~\cite{Lany_FERE_2008} for our dataset (quasi-\FERE)~\cite{Friedrich_CCE_2019} is also included.
    The red lines at $\pm$50 meV/atom indicate the typical \MAE\ of previous correction schemes~\cite{Jain_GGAU_PRB_2011,Stevanovic_FERE_2012}.
     \textcolor{black}{Adapted from}
    Ref.~\onlinecite{Friedrich_CCE_2019}.
    }
	\label{fig2}
\end{figure*}

The results from the binaries are then used to obtain the \CCE\ corrections and to apply them to the ternary test set.
Fig.~\ref{fig2} shows the example of the corrected \SCAN\ results where we also compare to an implementation of the \FERE\ method for our dataset (quasi-\FERE).
Both methods significantly decrease the plain \DFT\ errors by a factor of 2 to 4 with the \CCE\ \MAE\ of 27~meV/atom being about half of the quasi-\FERE\ one of 44~meV/atom.
The latter accuracy agrees well with the previously reported \MAEs\ of about 50~meV/atom of the other correction schemes~\cite{Jain_GGAU_PRB_2011,Stevanovic_FERE_2012}.
The \CCE\ errorbar is thus on the order of room temperature thermal energy and may as such enable more exact thermodynamic stability predictions.
\CCE\ achieves the same accuracy for ternary halides~\cite{Friedrich_CCE_2019}.
Since it is based on the bonding topology in materials, the method can also correct the relative stability of different polymorphs as demonstrated for several systems and can yield accurate defect energies as exemplified for Ti-O Magn{\'e}li phases~\cite{Friedrich_CCE_2019}.
For a few systems with large errors after correction in Fig.~\ref{fig2} for both methods (\CCE\ and quasi-\FERE) such as FeAl$_2$O$_4$, the experimental values might be inaccurate and should potentially be revised.

\begin{figure}[ht!]
	\centering
	\includegraphics[width=\columnwidth]{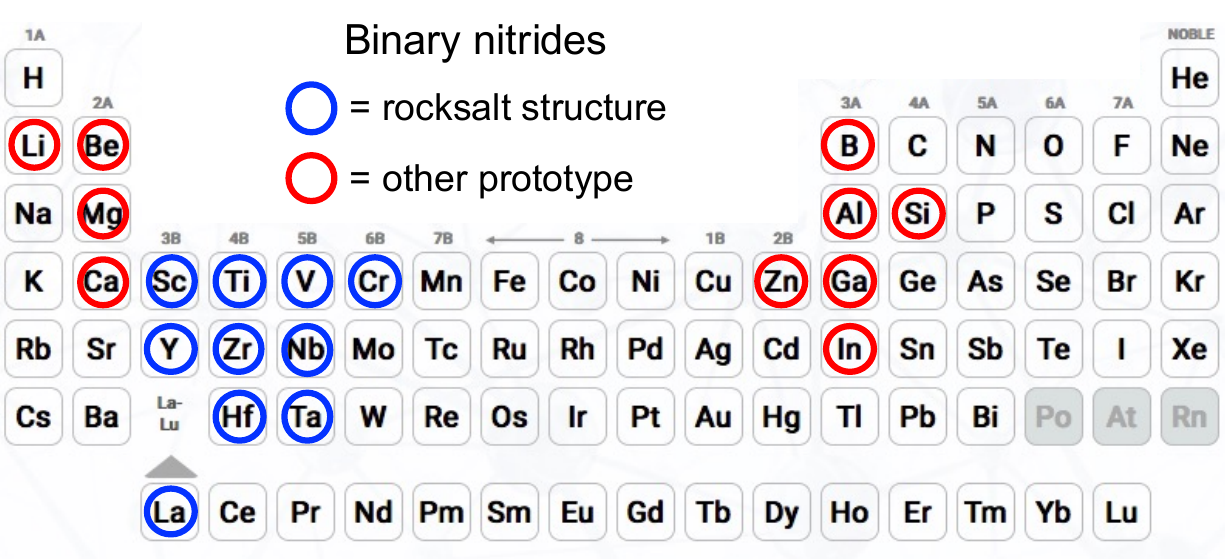}
	\caption{\small \textbf{Phase space of binary nitrides.}
	Periodic table with elements forming binary nitrides for which validated thermochemical data are available in standard collections~\cite{Kubaschewski_MTC_1993,Chase_NIST_JANAF_thermochem_tables_1998,Barin_1995,Wagman_NBS_thermodyn_tables_1982} highlighted by circles.
	Blue indicates the formation of rock salt structures and red stands for other structural prototypes.
    }
	\label{fig3}
\end{figure}

\noindent
{\bf Implementation: corrections for nitrides.}
The study of a suitable set of binary materials will first allow uncovering the errors of uncorrected \DFT\ for this materials class, as well as deriving the corrections.
The predictive power of the corrections will then be validated for ternary nitrides.
The computed and corrected data of this part are included in the Appendix.

For an overview, in Fig.~\ref{fig3}, elements forming binary nitrogen compounds with tabulated high quality thermochemical data in the standard collections~\cite{Kubaschewski_MTC_1993,Chase_NIST_JANAF_thermochem_tables_1998,Barin_1995,Wagman_NBS_thermodyn_tables_1982} are circled.
Nitrides are much more scarce than oxides since there are only 20 binaries.
The major reason is that the N$_2$ molecular reference phase is thermodynamically exceptionally stable which makes it difficult to form a nitrogen compound with a net energy gain.
Recent computational efforts have therefore also been devoted to predict metastable nitrides from suitable reactive nitrogen precursors~\cite{Sun_CoM_2017} and to build large stability maps of the uncharted nitrides materials space~\cite{Sun_NMAT_2019}.
We note that in the set of known binaries in Fig.~\ref{fig3} there is a predominance of rock salt structures indicated in blue as all $d$-element nitrides except Zn$_3$N$_2$ crystallize in this prototype.
These are expected to be very suitable precursors for high-entropy systems since many high-entropy carbides and oxides have this structure~\cite{curtarolo:art140,Oses_NatureReview_2020}.

\begin{figure*}[ht!]
	\centering
	\includegraphics[width=\textwidth]{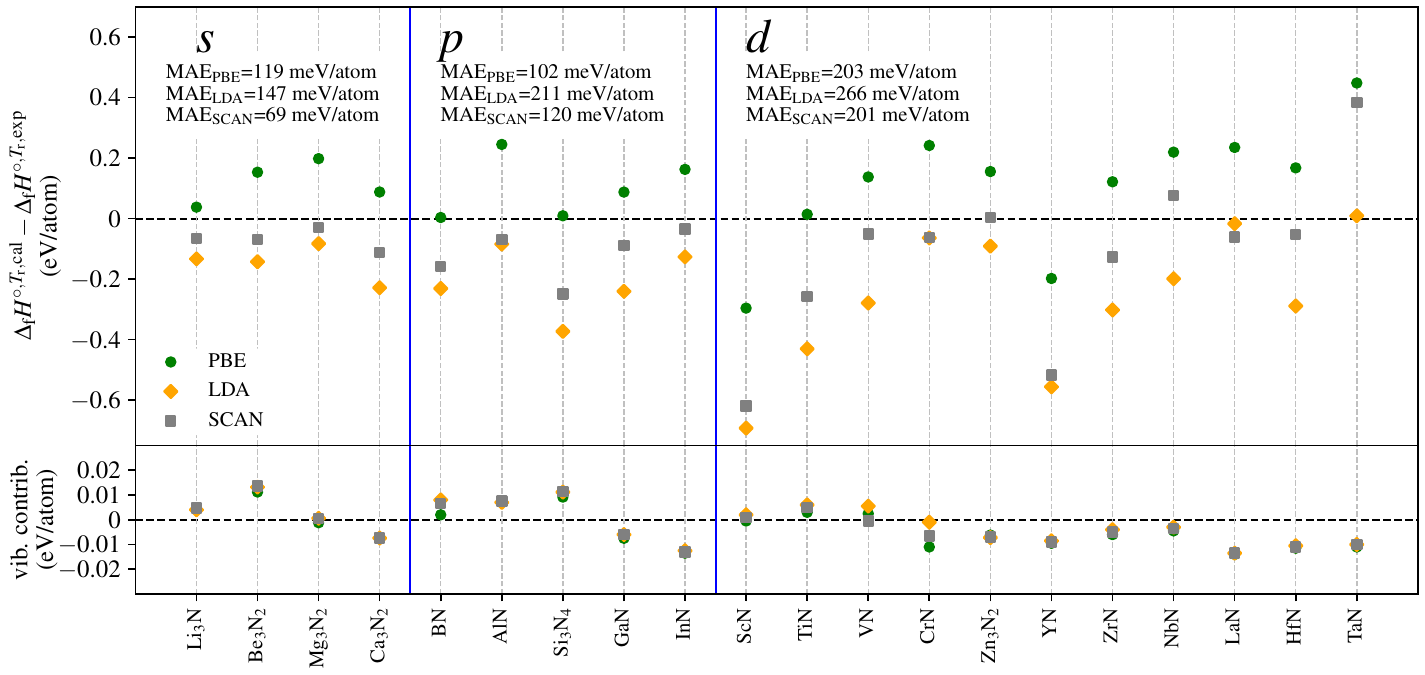}
	\caption{\small \textbf{Formation enthalpies and vibrational contribution for binary nitrides.}
	Differences between calculated and experimental room temperature formation enthalpies of 20 binary nitrides (upper panel) and vibrational (zero-point + thermal) contribution to the calculated formation enthalpy (lower panel) for three standard \DFT\ functionals.
    Vertical blue lines separate blocks with different $l$-character of the cations.
        }
	\label{fig4}
\end{figure*}

Figure~\ref{fig4} shows the difference between calculated \DFT+\AGL\ and experimental room temperature formation enthalpies, as well as the vibrational contribution in the lower panel.
The \MAEs\ for the nitrides are generally larger as for the oxides with \PBE\ again mostly underestimating the enthalpies with respect to experiment while \LDA\ and \SCAN\ typically lead to an overestimation.
Although there has been some debate about whether the tabulated experimental $\Delta_{\mathrm{f}}H$ for GaN~\cite{Kubaschewski_MTC_1993} should be updated with a more negative value~\cite{Lany_FERE_2008,Ranade_JPCB_2000}, we stick to the old result since only in this case it fits to the overall behavior that \LDA\ and \SCAN\ overestimate the formation enthalpy whereas \PBE\ does the opposite as observed for BN, AlN, and InN.
With the small vibrational contribution being at max of the order of 10-20~meV/atom with no significant differences between the approximations, the main error arises once more from the internal (\DFT) energy contribution.
Over the whole set, the three functionals exhibit significant \MAEs\ of 161, 229, and 155~meV/atom for \PBE, \LDA, and \SCAN, respectively outlining again the meta-\GGA\ as most accurate.
There is nevertheless a strong dependence on the $l$-character of the cation species.
While for the $s$-element nitrides, \SCAN\ exhibits the by far smallest \MAE\ of 69~meV/atom, \PBE\ performs best for the $p$-compounds and yields about the same average error as \SCAN\ for the $d$-systems.
\LDA\ shows the largest errors for all three groups in strict contrast to the oxides where it is significantly better than \PBE\ and, for the ternaries, even close to \SCAN.
Thus, this success of \LDA\ is not systematic for all materials classes.
It should be noted that, compared to the oxides, where the experimental errorbar is likely on the order of 10-20~meV/atom~\cite{Friedrich_CCE_2019}, the average error of the Kubaschewski values~\cite{Kubaschewski_MTC_1993} given for all binaries except TiN is $\sim$33~meV/atom.

\onecolumngrid

\LTcapwidth=\textwidth

\newcommand\hdl[1]{\multicolumn{1}{l}{#1}}
\newcommand\hdr[1]{\multicolumn{1}{r}{#1}}
\newcommand\hd[1]{\multicolumn{1}{c}{#1}}
\newcommand\zz[1]{^{#1}\!\!}
\setlength\tabcolsep{11pt}
\begin{longtable*}{@{}lr|*{2}{D..{\tablecolumnspacing}}|*{2}{D..{\tablecolumnspacing}}|*{2}{D..{\tablecolumnspacing}}|*{1}{D..{\tablecolumnspacing}}@{}}
  \caption{\small \textbf{\CCE\ corrections for 298.15 and 0~K for nitrides.}
    Corrections per bond $\delta H^{T,A^{+\alpha}}_{A-Y}$ of the \CCE\ method for each cation species $A$ in oxidation states $+\alpha$ for 298.15 and 0~K obtained from binary nitrides for the different functionals.
    The experimental formation enthalpies per bond that can be used to obtain a rough estimate of the $\Delta_{\mathrm{f}} H$ value of a material without \DFT\ calculations according to Ref.~\onlinecite{Friedrich_CCE_2019,Friedrich_PRM_2021}, are given in the last column.
    All corrections are in eV/bond.
  }\label{tab_1}\\
  \hline
  cation & \hd{$+\alpha$} & \multicolumn{2}{c}{\PBE} & \multicolumn{2}{c}{\LDA} & \multicolumn{2}{c}{\SCAN} & \multicolumn{1}{c}{\CCE@exp} \\
  species $A$ & & \hd{298.15 K} & \hd{0 K} & \hd{298.15 K} & \hd{0 K} & \hd{298.15 K} & \hd{0 K} & \hd{298.15 K} \\
  \hline
  \endfirsthead

  \multicolumn{9}{c}
  {{\tablename\ \thetable{}. (\textit{continued})}} \\
  \hline
  cation & \hd{$+\alpha$} & \multicolumn{2}{c}{\PBE} & \multicolumn{2}{c}{\LDA} & \multicolumn{2}{c}{\SCAN} & \multicolumn{1}{c}{\CCE@exp} \\
  species $A$ & & \hd{298.15 K} & \hd{0 K} & \hd{298.15 K} & \hd{0 K} & \hd{298.15 K} & \hd{0 K} & \hd{298.15 K} \\
  \hline
	\endhead
Li    &     +1     &     0.01663      &      0.00912    &  -0.06850      &     -0.07562    &  -0.03588       &     -0.04338    &       -0.21350     \\
Be    &     +2     &     0.05908      &      0.05292    &  -0.06500      &     -0.07133    &  -0.03433       &     -0.04075    &       -0.50917     \\
B     &     +3     &     0.00100      &     -0.00867    &  -0.15933      &     -0.17000    &  -0.11000       &     -0.12067    &       -0.86700     \\
Mg    &     +2     &     0.08300      &      0.07642    &  -0.03475      &     -0.04167    &  -0.01242       &     -0.01950    &       -0.39858     \\
Al    &     +3     &     0.11875      &      0.10850    &  -0.04575      &     -0.05575    &  -0.03850       &     -0.04875    &       -0.82500     \\
Si    &     +4     &     0.00000      &     -0.01300    &  -0.22408      &     -0.23733    &  -0.15183       &     -0.16508    &       -0.64325     \\
Ca    &     +2     &     0.03967      &      0.03400    &  -0.09225      &     -0.09792    &  -0.04392       &     -0.04942    &       -0.37225     \\
Sc    &     +3     &    -0.09850      &     -0.10500    &  -0.23150      &     -0.23833    &  -0.20667       &     -0.21350    &       -0.54200     \\
Ti    &     +3     &     0.00367      &     -0.00283    &  -0.14550      &     -0.15233    &  -0.08733       &     -0.09400    &       -0.58400     \\
V     &     +3     &     0.04500      &      0.03917    &  -0.09483      &     -0.10100    &  -0.01683       &     -0.02250    &       -0.37650     \\
Cr    &     +3     &     0.08417      &      0.08167    &  -0.02100      &     -0.02583    &  -0.01867       &     -0.02233    &       -0.20250     \\
Zn    &     +2     &     0.06733      &      0.06225    &  -0.03492      &     -0.03908    &   0.00417       &     -0.00025    &       -0.01950     \\
Ga    &     +3     &     0.04725      &      0.03950    &  -0.11725      &     -0.12550    &  -0.04200       &     -0.05075    &       -0.28400     \\
Y     &     +3     &    -0.06283      &     -0.06750    &  -0.18267      &     -0.18783    &  -0.16950       &     -0.17450    &       -0.51683     \\
Zr    &     +3     &     0.04250      &      0.03733    &  -0.09933      &     -0.10500    &  -0.04067       &     -0.04617    &       -0.63100     \\
Nb    &     +3     &     0.07467      &      0.06967    &  -0.06550      &     -0.07083    &   0.02667       &      0.02150    &       -0.40833     \\
In    &     +3     &     0.08825      &      0.08350    &  -0.05700      &     -0.06225    &  -0.01125       &     -0.01650    &       -0.04450     \\
La    &     +3     &     0.08300      &      0.07983    &  -0.00100      &     -0.00433    &  -0.01600       &     -0.01950    &       -0.52400     \\
Hf    &     +3     &     0.05983      &      0.05633    &  -0.09283      &     -0.09683    &  -0.01433       &     -0.01817    &       -0.64533     \\
Ta    &     +3     &     0.15283      &      0.14933    &   0.00617      &      0.00233    &   0.13100       &      0.12733    &       -0.43583     \\
	\hline
\end{longtable*}

\twocolumngrid

The results from the binaries are then used to obtain the \CCE\ corrections for nitrides which are listed in Table~\ref{tab_1} including also the experimental formation enthalpies per bond \CCE@exp~\cite{Friedrich_CCE_2019,Friedrich_PRM_2021}.
As shown by us previously in Refs.~\onlinecite{Friedrich_CCE_2019,Friedrich_PRM_2021}, the vibrational contribution does not need to be taken into account explicitly and the corrections can be directly fitted to the plain \DFT\ formation enthalpies.
This gives two types of corrections: (i) for room temperature (298.15 K) and (ii) for 0 K, meaning that when they are applied to the plain \DFT\ enthalpies of target compounds, they yield corrected estimates for the respective temperature.

\begin{figure}[ht!]
	\centering
	\includegraphics[width=\columnwidth]{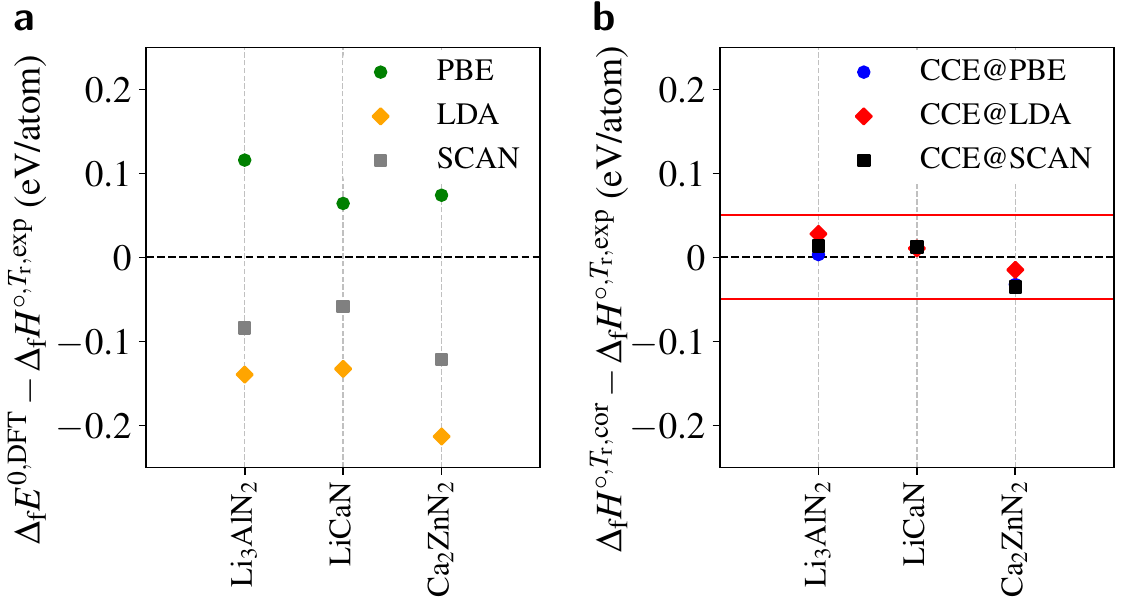}
	\caption{\small \textbf{Uncorrected and corrected enthalpies for ternary nitrides.}
	Differences between calculated (\textbf{a}) as well as corrected (\textbf{b}) and experimental room temperature formation enthalpies for seven ternary nitrides.
	The red lines at $\pm$50 meV/atom indicate the typical \MAE\ of previous correction schemes~\cite{Jain_GGAU_PRB_2011,Stevanovic_FERE_2012}.
        }
	\label{fig5}
\end{figure}

The amount of ternary nitrides in particular with validated thermochemical and structural data which are also available in the \AFLOW-{\small ICSD} database is, however, very scarce, with no entries in the standard collections that were used for the oxides.
In Fig.~\ref{fig5} we present both the uncorrected and corrected results for three compounds with measured formation enthalpies reported in Refs.~\cite{McHale_ChemMater_1997,McHale_ChemMater_1999}.
As expected, the plain \DFT\ results show large errors for all three functionals (\MAEs\ of 85, 162, and 88 meV/atom for \PBE, \LDA, and \SCAN).
When \CCE\ is applied, the \MAEs\ reduce to 16, 18, and 21~meV/atom again confirming the high accuracy of \CCE\ enthalpies although of course average values for a set of only three entries must be taken with some care.

\vspace{0.5cm}
The oxide and nitride corrections are available from the \AFLOW-\CCE\ module which automatically determines the oxidation numbers of all ions and only needs a structure as input~\cite{Friedrich_PRM_2021}.
It has been integrated into the \AFLOW\ software~\cite{Oses_CMS_2023}
of version 3.2.7 or later for which the source code is available at
\href{http://aflow.org/install-aflow/}{http://aflow.org/install-aflow/}
and
\href{http://materials.duke.edu/AFLOW/}{http://materials.duke.edu/AFLOW/}.
The implementation also includes a Python environment that is distributed with the \AFLOW\ source and can be generated with the command \texttt{aflow --cce
--print=python}.
The \CCE\ web tool~\cite{Esters_CMS_2023} can be accessed via:
\href{http://aflow.org/aflow-online/}{http://aflow.org/aflow-online/}.
The main commands for the \AFLOW-\CCE\ command line tool are summarized in Table~\ref{tab_2}.
Instructions are available through the \AFLOW-School:
\href{http://aflow.org/aflow-school/}{http://aflow.org/aflow-school/}.
All options and further details are described in Ref.~\cite{Friedrich_PRM_2021}.

    \begin{table*}[ht!]
        \caption{\small \textbf{} Main commands for the \AFLOW-\CCE\ command line tool.
        }\label{tab_2}
        \begin{tabular}{p{0.45\textwidth} |p{0.45\textwidth}}
            \hline
            Command & Description \\
            \hline
            \texttt{aflow --cce} & Prints user instructions. \\
            \texttt{aflow --cce={\small STRUCTURE\_FILE\_PATH}}  & Provides the output of the full \CCE\ analysis, \emph{i.e.} cation coordination numbers, oxidation numbers, and \CCE\ corrections and formation enthalpies, for the structure in {\small STRUCTURE\_FILE\_PATH}. \\
            \texttt{aflow --get\_cce\_corrections < {\small STRUCTURE\_FILE\_PATH}} & Returns the \CCE\ corrections and formation enthalpies for the structure in {\small STRUCTURE\_FILE\_PATH}. \\
            \texttt{aflow --get\_oxidation\_number < {\small STRUCTURE\_FILE\_PATH}} & Gives the oxidation numbers for the structure in {\small STRUCTURE\_FILE\_PATH}. \\
            \texttt{aflow --get\_cation\_coord\_num < {\small STRUCTURE\_FILE\_PATH}} & Determines the cation coordination numbers for the structure in {\small STRUCTURE\_FILE\_PATH}. \\
            \hline
        \end{tabular}
    \end{table*}

\noindent
{\bf Future implementations.}
The \CCE\ parameterization will be extended to all other relevant anion classes.
These include: (i) all halides, \emph{i.e.} fluorides, chlorides, bromides, and iodides; (ii) all additional chalcogenides, \emph{i.e.} sulfides, selenides, and tellurides; (iii) additional pnictides, \emph{i.e.} phosphides, arsenides, and potentially antimonides; and (iv) possibly hydrides.
The method will also be leveraged for the challenging design of disordered high-entropy ceramics by providing an ensemble of corrections for the partial occupation (\AFLOW-{\small POCC})~\cite{curtarolo:art110} algorithm used to model these materials.
It is also expected to become very useful for energetic corrections of nanostructures and in surface science based on the bonding topology, where for instance, the relative stability of different quasi 2D systems and with respect to bulk materials~\cite{Puthirath_Balan_NNANO_2018,Friedrich_NanoLett_2022,Barnowsky_AdvElMats_2023,Balan_MatTod_2022,Kaur_AdvMat_2022} must be estimated accurately.
The temperature dependence can be implemented in a dynamic fashion allowing for enthalpy estimates at any given temperature on the fly.

\ \\
\section*{Conclusions} \label{conclusions}
\noindent
We have presented the computational capabilities of the \AFLOW-\CCE\ software for thermodynamic stability predictions utilizing our previous oxide data and new results for nitrides.
When calculating the formation enthalpies from plain \DFT\ for several standard functionals such as \LDA, \PBE, and \SCAN, average errors compared to high-quality experimental data of the order of several hundred meV/atom are found.
For oxides, \SCAN\ shows the best performance achieving $\sim$100~meV/atom closely followed by \LDA.
Also for nitrides, \SCAN\ depicts the smallest mean errors slightly above 150~meV/atom almost matched by \PBE\ with \LDA\ revealing larger deviations.

Since the accuracy of these approximations is too low for materials design at room temperature ($\sim$25~meV), we employ the \AFLOW-\CCE\ implementation of our previously developed coordination corrected enthalpies method for the accurate evaluation of the thermodynamic stability of ionic materials.
\CCE\ is based on an intuitive parametrization of \DFT\ errors with respect to coordination numbers and cation oxidation states.
The method has been benchmarked for the formation enthalpies of oxides and nitrides revealing mean errors on the order of $\sim$25~meV/atom.
In addition, it can rectify errors in relative stability predictions of polymorphs from plain \DFT.
The corrections are freely available in the \AFLOW-\CCE\ module for the automatic correction of enthalpies based on only structural inputs.
The method and its implementation can be valuable for correcting the results in large materials databases to enable advanced materials predictions.

\section*{Acknowledgments}

\noindent
The authors thank David Hicks, Cormac Toher, Ohad Levy, Simon Divilov, Hagen Eckert, Michael Mehl, Adam Zettel, Corey Oses, Arrigo Calzolari, Arkady V. Krasheninnikov, Thomas Heine, and Xiomara Campilongo for fruitful discussions.
S.C. acknowledges the DoD SPICES MURI
sponsored by the Office of Naval Research (Naval Research contract
N00014-21-1-2515) for financial support.
R.F. acknowledges funding for the “Autonomous Materials Thermodynamics” (AutoMaT) project by Technische Universität Dresden and Helmholtz-Zentrum Dresden-Rossendorf within the DRESDEN-concept alliance.
R.F. acknowledges support from the Alexander von Humboldt foundation under the Feodor Lynen research fellowship.
The authors thank the HZDR Computing Center, HLRS Stuttgart (HAWK cluster), the Paderborn Center for Parallel Computing (PC2, Noctua 2 cluster),
and the DoD High Performance Computing Modernization Program for computational support.

\appendix

\onecolumngrid

\section{Tables with numerical data}

\def\protogeneric{{http://www.aflowlib.org/prototype-encyclopedia/prototype_index.html}}

\LTcapwidth=\textwidth
\setlength\tabcolsep{4pt}
\begin{longtable*}{@{}l|cccc@{}}
	\caption{\small {\bf Structural data for binary and ternary nitrides.}
	\ICSD\  numbers, space group numbers, Pearson symbols, and \AFLOW\ prototype labels~\cite{curtarolo:art121,curtarolo:art145,curtarolo:art173} for the 20 binary and three ternary nitrides.
			Space-groups and Pearson symbols are calculated with \AFLOWSYM~\cite{curtarolo:art135}. Note: the order of the Wyckoff position letters in the prototype column follows the alphabetic syntax to generate the prototype, while the associated web-link points to the standardized description in the \AFLOW\ Encyclopedia of Crystallographic Prototypes~\citePROTOS.}
	\label{tab_app2_1}\\
	\hline
	formula  & \ICSD\ \# & space group \# & Pearson symbol & \AFLOW\ prototype \\
	\hline
	\endfirsthead
	\multicolumn{5}{c}
	{{\tablename\ \thetable{}. (\textit{continued})}} \\
	\hline
	formula  & \ICSD\ \# & space group \# & Pearson symbol & \AFLOW\ prototype \\
	\hline
	\endhead
	Li$_3$N		  & 76944  & 191 & hP4  & \href{http://aflow.org/prototype-encyclopedia/A3B_hP4_191_bc_a.html}{A3B\_hP4\_191\_bc\_a}   \\
	Be$_3$N$_2$	  & 616348 & 206 & cI80 & \href{\protogeneric}{A3B2\_cI80\_206\_e\_bd} \\
	Mg$_3$N$_2$	  & 411210 & 206 & cI80 & \href{\protogeneric}{A3B2\_cI80\_206\_e\_bd} \\
	Ca$_3$N$_2$	  & 50991  & 206 & cI80 & \href{\protogeneric}{A3B2\_cI80\_206\_e\_bd} \\
	BN            & 186246 & 194 & hP4  & \href{http://aflow.org/prototype-encyclopedia/AB_hP4_194_c_d.html}{AB\_hP4\_194\_d\_c}    \\
	AlN	          & 602459 & 186 & hP4  & \href{http://aflow.org/prototype-encyclopedia/AB_hP4_186_b_b.html}{AB\_hP4\_186\_b\_b}     \\
	Si$_3$N$_4$	  & 74744  & 176 & hP14 & \href{http://aflow.org/prototype-encyclopedia/A4B3_hP14_176_ch_h.html}{A4B3\_hP14\_176\_ch\_h} \\
	GaN	          & 156259 & 186 & hP4  & \href{http://aflow.org/prototype-encyclopedia/AB_hP4_186_b_b.html}{AB\_hP4\_186\_b\_b}     \\
	InN	          & 157515 & 186 & hP4  & \href{http://aflow.org/prototype-encyclopedia/AB_hP4_186_b_b.html}{AB\_hP4\_186\_b\_b}     \\
	ScN	          & 26948  & 225 & cF8  & \href{http://aflow.org/prototype-encyclopedia/AB_cF8_225_a_b.html}{AB\_cF8\_225\_b\_a}     \\
	TiN           & 64907  & 225 & cF8  & \href{http://aflow.org/prototype-encyclopedia/AB_cF8_225_a_b.html}{AB\_cF8\_225\_b\_a}     \\
	VN            & 76526  & 225 & cF8  & \href{http://aflow.org/prototype-encyclopedia/AB_cF8_225_a_b.html}{AB\_cF8\_225\_b\_a}     \\
	CrN           & 41827  & 225 & cF8  & \href{http://aflow.org/prototype-encyclopedia/AB_cF8_225_a_b.html}{AB\_cF8\_225\_a\_b}     \\
	Zn$_3$N$_2$	  & 84918  & 206 & cI80 & \href{\protogeneric}{A2B3\_cI80\_206\_ad\_e} \\
	YN            & 76528  & 225 & cF8  & \href{http://aflow.org/prototype-encyclopedia/AB_cF8_225_a_b.html}{AB\_cF8\_225\_b\_a}     \\
	ZrN           & 167851 & 225 & cF8  & \href{http://aflow.org/prototype-encyclopedia/AB_cF8_225_a_b.html}{AB\_cF8\_225\_a\_b}     \\
	NbN           & 183423 & 225 & cF8  & \href{http://aflow.org/prototype-encyclopedia/AB_cF8_225_a_b.html}{AB\_cF8\_225\_b\_a}     \\
	LaN           & 641462 & 225 & cF8  & \href{http://aflow.org/prototype-encyclopedia/AB_cF8_225_a_b.html}{AB\_cF8\_225\_a\_b}     \\
	HfN           & 53025  & 225 & cF8  & \href{http://aflow.org/prototype-encyclopedia/AB_cF8_225_a_b.html}{AB\_cF8\_225\_a\_b}     \\
	TaN           & 644727 & 225 & cF8  & \href{http://aflow.org/prototype-encyclopedia/AB_cF8_225_a_b.html}{AB\_cF8\_225\_b\_a}     \\
	\hline
	Li$_3$AlN$_2$ & 25565  & 206 & cI96 & \href{http://aflow.org/prototype-encyclopedia/AB3C2_cI96_206_c_e_ad.html}{AB3C2\_cI96\_206\_c\_e\_bd} \\
	LiCaN         & 107304 & 62 & oP12 & \href{http://aflow.org/prototype-encyclopedia/ABC_oP12_62_c_c_c.html}{ABC\_oP12\_62\_c\_c\_c} \\
	Ca$_2$ZnN$_2$  & 69049  & 139 & tI10 & \href{\protogeneric}{A2B2C\_tI10\_139\_e\_e\_a} \\
	\hline
\end{longtable*}

\newpage

{
\LTcapwidth=\textwidth
\setlength\tabcolsep{2pt}
\begin{longtable*}{@{}l|ccccccccc@{}}
	\caption{\small {\bf Formation enthalpies for binary and ternary nitrides.}
	Calculated \DFT+\AGL, plain \DFT\ and experimental room temperature formation enthalpies, number of cation-anion bonds $N_{A-Y}$ per formula unit and oxidation states $+\alpha$ of the cation(s) for binary and ternary nitrides.
	In case of ternaries, for the cation-nitrogen bonds- and oxidation numbers, the first (second) number in the column refers to the first (second) element in the formula.
		Experimental values are from Kubaschewski~\emph{et al.}~\cite{Kubaschewski_MTC_1993}, \NISTJANAF~\cite{Chase_NIST_JANAF_thermochem_tables_1998}, Barin~\cite{Barin_1995}, and McHale~\emph{et al.}~\cite{McHale_ChemMater_1997,McHale_ChemMater_1999}.
		Enthalpies are in eV/atom; corrections are in eV/bond.}\label{tab_app3_1}\\
	\hline
	formula & \hd{\PBE+\AGL} & \hd{\PBE} & \hd{\LDA+\AGL} & \hd{\LDA} & \hd{\SCAN+\AGL} & \hd{\SCAN} & \hd{Exp.} & \hd{$N_{A-Y}$} & \hd{$+\alpha$} \\
	\hline
	\endfirsthead
	\multicolumn{10}{c}
	{{\tablename\ \thetable{}. (\textit{continued})}} \\
	\hline
	formula & \hd{\PBE+\AGL} & \hd{\PBE} & \hd{\LDA+\AGL} & \hd{\LDA} & \hd{\SCAN+\AGL} & \hd{\SCAN} & \hd{Exp.} & \hd{$N_{A-Y}$} & \hd{$+\alpha$} \\
	\hline
	\endhead
	Li$_3$N		  & $-$0.389 & $-$0.394 & $-$0.560 & $-$0.564 & $-$0.494 & $-$0.499 & $-$0.427~\cite{Kubaschewski_MTC_1993}                   & 8  & +1 \\
	Be$_3$N$_2$	  & $-$1.069 & $-$1.080 & $-$1.365 & $-$1.378 & $-$1.291 & $-$1.304 & $-$1.222~\cite{Kubaschewski_MTC_1993}                   & 12 & +2 \\
	Mg$_3$N$_2$	  & $-$0.759 & $-$0.757 & $-$1.039 & $-$1.040 & $-$0.986 & $-$0.986 & $-$0.957~\cite{Kubaschewski_MTC_1993}                   & 12 & +2 \\
	Ca$_3$N$_2$	  & $-$0.806 & $-$0.798 & $-$1.122 & $-$1.115 & $-$1.006 & $-$0.999 & $-$0.893~\cite{Barin_1995}                              & 12 & +2 \\
	BN            & $-$1.297 & $-$1.299 & $-$1.532 & $-$1.539 & $-$1.459 & $-$1.466 & $-$1.300~\cite{Chase_NIST_JANAF_thermochem_tables_1998} & 3  & +3 \\
	AlN	          & $-$1.405 & $-$1.413 & $-$1.735 & $-$1.742 & $-$1.719 & $-$1.727 & $-$1.650~\cite{Kubaschewski_MTC_1993}                   & 4  & +3 \\
	Si$_3$N$_4$	  & $-$1.094 & $-$1.103 & $-$1.476 & $-$1.487 & $-$1.352 & $-$1.363 & $-$1.103~\cite{Kubaschewski_MTC_1993}                   & 12 & +4 \\
	GaN	          & $-$0.481 & $-$0.473 & $-$0.808 & $-$0.802 & $-$0.658 & $-$0.652 & $-$0.568~\cite{Kubaschewski_MTC_1993}                   & 4  & +3 \\
	InN	          & 0.074    & 0.087    & $-$0.215 & $-$0.203 & $-$0.124 & $-$0.111 & $-$0.089~\cite{Barin_1995}                              & 4  & +3 \\
	ScN	          & $-$1.922 & $-$1.922 & $-$2.318 & $-$2.320 & $-$2.245 & $-$2.246 & $-$1.626~\cite{Kubaschewski_MTC_1993}                   & 6  & +3 \\
	TiN           & $-$1.738 & $-$1.741 & $-$2.182 & $-$2.188 & $-$2.009 & $-$2.014 & $-$1.752~\cite{Kubaschewski_MTC_1993}                   & 6  & +3 \\
	VN            & $-$0.992 & $-$0.994 & $-$1.408 & $-$1.414 & $-$1.180 & $-$1.180 & $-$1.129~\cite{Kubaschewski_MTC_1993}                   & 6  & +3 \\
	CrN           & $-$0.366 & $-$0.355 & $-$0.672 & $-$0.670 & $-$0.670 & $-$0.663 & $-$0.608~\cite{Kubaschewski_MTC_1993}                   & 6  & +3 \\
	Zn$_3$N$_2$	  & 0.109    & 0.115    & $-$0.138 & $-$0.131 & $-$0.044 & $-$0.037 & $-$0.047~\cite{Kubaschewski_MTC_1993}                   & 12 & +2 \\
	YN            & $-$1.749 & $-$1.739 & $-$2.107 & $-$2.098 & $-$2.068 & $-$2.059 & $-$1.550~\cite{Kubaschewski_MTC_1993}                   & 6  & +3 \\
	ZrN           & $-$1.771 & $-$1.765 & $-$2.195 & $-$2.191 & $-$2.020 & $-$2.015 & $-$1.893~\cite{Chase_NIST_JANAF_thermochem_tables_1998} & 6  & +3 \\
	NbN           & $-$1.006 & $-$1.001 & $-$1.424 & $-$1.421 & $-$1.149 & $-$1.145 & $-$1.225~\cite{Kubaschewski_MTC_1993}                   & 6  & +3 \\
	LaN           & $-$1.337 & $-$1.323 & $-$1.589 & $-$1.575 & $-$1.634 & $-$1.620 & $-$1.572~\cite{Barin_1995}                              & 6  & +3 \\
	HfN           & $-$1.768 & $-$1.757 & $-$2.225 & $-$2.215 & $-$1.990 & $-$1.979 & $-$1.936~\cite{Kubaschewski_MTC_1993}                   & 6  & +3 \\
	TaN           & $-$0.860 & $-$0.849 & $-$1.299 & $-$1.289 & $-$0.924 & $-$0.914 & $-$1.308~\cite{Kubaschewski_MTC_1993}                   & 6  & +3 \\
	\hline
	Li$_3$AlN$_2$ & $-$0.853 & $-$0.865 & $-$1.109 & $-$1.120 & $-$1.053 & $-$1.065 & $-$0.981~\cite{McHale_ChemMater_1999}                   & 12, 4  & +1, +3 \\
	LiCaN         & $-$0.667 & $-$0.663 & $-$0.929 & $-$0.926 & $-$0.830 & $-$0.827 & $-$0.562~\cite{McHale_ChemMater_1997}                   & 3, 4 & +1, +2   \\
	Ca$_2$ZnN$_2$ & $-$0.718 & $-$0.711 & $-$1.006 & $-$0.999 & $-$0.914 & $-$0.907 & $-$0.785~\cite{McHale_ChemMater_1997}                   & 10, 2  & +2, +2 \\
	\hline
\end{longtable*}
}

{
  \LTcapwidth=\textwidth
  \begin{longtable*}{@{}l|*{1}{D..{2.3}}*{1}{D..{2.3}}*{1}{D..{2.3}}|*{1}{D..{2.3}}*{1}{D..{2.3}}*{1}{D..{2.3}}|*{1}{D..{2.3}}*{1}{D..{2.3}}*{1}{D..{2.3}}@{}}
    \caption{\small {\bf \AGL\ contributions to the formation enthalpies for binary and ternary nitrides.}
      Total vibrational (\TVC), zero-point (\ZPC) and thermal (\TC) contributions to the calculated formation enthalpies obtained from \AGL~\cite{Blanco_jmolstrthch_1996,BlancoGIBBS2004,curtarolo:art96,curtarolo:art115,Poirier_Earth_Interior_2000} for each functional for binary and ternary nitrides.
      The sum of \ZPC\ and \TC\ might not match exactly the total \AGL\ contribution listed due to rounding.
      All values are in eV/atom.}\label{tab_app6_1}\\
    \hline
    formula & \multicolumn{3}{c|}{\PBE} & \multicolumn{3}{c|}{\LDA} & \multicolumn{3}{c}{\SCAN} \\
    & \hd{\AGL-\TVC} & \hd{\AGL-\ZPC} & \multicolumn{1}{c|}{\AGL-\TC}  & \hd{\AGL-\TVC} & \hd{\AGL-\ZPC} & \multicolumn{1}{c|}{\AGL-\TC} & \hd{\AGL-\TVC} & \hd{\AGL-\ZPC} & \hd{\AGL-\TC} \\
    \hline
    \endfirsthead
    \multicolumn{10}{c}
    {{\tablename\ \thetable{}. (\textit{continued})}} \\
    \hline
    formula & \multicolumn{3}{c|}{\PBE} & \multicolumn{3}{c|}{\LDA} & \multicolumn{3}{c}{\SCAN} \\
    & \hd{\AGL-\TVC} & \hd{\AGL-\ZPC} & \multicolumn{1}{c|}{\AGL-\TC}  & \hd{\AGL-\TVC} & \hd{\AGL-\ZPC} & \multicolumn{1}{c|}{\AGL-\TC} & \hd{\AGL-\TVC} & \hd{\AGL-\ZPC} & \hd{\AGL-\TC} \\
    \hline
    \endhead
    Li$_3$N		  &	 0.004  &	 0.019  &	-0.015 &  0.004  &	  0.018  &	 -0.014 &  0.005  &	 0.020  &	-0.015	\\
    Be$_3$N$_2$	  &	 0.011  &	 0.026  &	-0.015 &  0.013  &	  0.028  &	 -0.015 &  0.014  &	 0.029  &	-0.015	\\
    Mg$_3$N$_2$	  &	-0.001  &	 0.015  &	-0.016 &  0.001  &	  0.017  &	 -0.017 &  0.001  &	 0.017  &	-0.017	\\
    Ca$_3$N$_2$	  &	-0.007  &	 0.006  &	-0.014 & -0.007  &	  0.006  &	 -0.014 & -0.007  &	 0.006  &	-0.013	\\
    BN            &	 0.002  &	 0.017  &	-0.015 &  0.008  &	  0.024  &	 -0.016 &  0.007  &	 0.022  &	-0.016	\\
    AlN	          &	 0.007  &	 0.028  &	-0.021 &  0.007  &	  0.027  &	 -0.020 &  0.007  &	 0.028  &	-0.021	\\
    Si$_3$N$_4$	  &	 0.009  &	 0.031  &	-0.022 &  0.011  &	  0.034  &	 -0.023 &  0.011  &	 0.034  &	-0.023	\\
    GaN	          &	-0.007  &	 0.008  &	-0.016 & -0.006  &	  0.010  &	 -0.016 & -0.006  &	 0.012  &	-0.018	\\
    InN	          &	-0.014  &	-0.004  &	-0.009 & -0.013  &	 -0.002  &	 -0.010 & -0.013  &	-0.002  &	-0.011	\\
    ScN	          &	 0.000  &	 0.019  &	-0.020 &  0.002  &	  0.023  &	 -0.021 &  0.001  &	 0.021  &	-0.020	\\
    TiN           &	 0.003  &	 0.022  &	-0.019 &  0.006  &	  0.026  &	 -0.020 &  0.005  &	 0.025  &	-0.020	\\
    VN            &	 0.002  &	 0.020  &	-0.018 &  0.006  &	  0.024  &	 -0.019 &  0.000  &	 0.017  &	-0.017	\\
    CrN           &	-0.011  &	-0.004  &	-0.008 & -0.001  &	  0.014  &	 -0.015 & -0.006  &	 0.005  &	-0.011	\\
    Zn$_3$N$_2$	  &	-0.006  &	 0.006  &	-0.012 & -0.007  &	  0.003  &	 -0.010 & -0.007  &	 0.004  &	-0.011	\\
    YN            &	-0.010  &	 0.004  &	-0.014 & -0.008  &	  0.007  &	 -0.015 & -0.009  &	 0.006  &	-0.015	\\
    ZrN           &	-0.006  &	 0.010  &	-0.016 & -0.004  &	  0.013  &	 -0.017 & -0.005  &	 0.011  &	-0.016	\\
    NbN           &	-0.005  &	 0.010  &	-0.015 & -0.003  &	  0.013  &	 -0.016 & -0.004  &	 0.012  &	-0.016	\\
    LaN           &	-0.014  &	-0.005  &	-0.009 & -0.014  &	 -0.003  &	 -0.010 & -0.014  &	-0.003  &	-0.011	\\
    HfN           &	-0.012  &	-0.001  &	-0.011 & -0.011  &	  0.001  &	 -0.012 & -0.011  &	 0.001  &	-0.012	\\
    TaN           &	-0.011  &	-0.001  &	-0.010 & -0.010  &	  0.002  &	 -0.011 & -0.010  &	 0.001  &	-0.011	\\
	\hline
    Li$_3$AlN$_2$ &  0.012  &    0.032  &   -0.021 &  0.011  &    0.031  &   -0.020 &  0.012  &  0.033  &   -0.021	\\
    LiCaN         & -0.003  &    0.010  &   -0.013 & -0.003  &    0.010  &   -0.013 & -0.003  &  0.010  &   -0.013	\\
    Ca$_2$ZnN$_2$ & -0.007  &    0.006  &   -0.013 & -0.007  &    0.005  &   -0.012 & -0.007  &  0.005  &   -0.012	\\
	\hline
\end{longtable*}
}

{
\LTcapwidth=\textwidth
\begin{longtable*}{@{}l|cccc}
	\caption{\small {\bf \CCE\ formation enthalpies for ternary nitrides.}
	\CCE\ formation enthalpies (using the corrections from Table~\ref{tab_1}) and experimental room temperature values for ternary nitrides.
	Experimental data are from McHale~\emph{et al.}~\cite{McHale_ChemMater_1997,McHale_ChemMater_1999}.
	All enthalpies are in eV/atom.}\label{tab_app9_1}\\
	\hline
	formula & \CCE@ & \CCE@ & \CCE@ & Exp. \\
	& \PBE\ & \LDA\ & \SCAN\ & \\
	\hline
	\endfirsthead
	\multicolumn{5}{c}
	{{\tablename\ \thetable{}. (\textit{continued})}} \\
	\hline
	formula & \PBE\ & \LDA\ & \SCAN\ & Exp. \\
	& \CCE\ & \CCE\ & \CCE\ & \\
	\hline
	\endhead
	Li$_3$AlN$_2$ & $-$0.978 & $-$0.953 & $-$0.967 & $-$0.981~\cite{McHale_ChemMater_1999} \\
	LiCaN         & $-$0.550 & $-$0.551 & $-$0.549 & $-$0.562~\cite{McHale_ChemMater_1997} \\
	Ca$_2$ZnN$_2$ & $-$0.818 & $-$0.800 & $-$0.821 & $-$0.785~\cite{McHale_ChemMater_1997} \\
	\hline
\end{longtable*}
}

\twocolumngrid

\newcommand{\Ozolins}{Ozoli{\c{n}}{\v{s}}}


\begin{thebibliography}{10}
\expandafter\ifx\csname urlstyle\endcsname\relax
  \providecommand{\doi}[1]{doi:\discretionary{}{}{}#1}\else
  \providecommand{\doi}{doi:\discretionary{}{}{}\begingroup
  \urlstyle{rm}\Url}\fi
\providecommand{\selectlanguage}[1]{\relax}
\providecommand{\bibAnnoteFile}[1]{%
  \IfFileExists{#1}{\begin{quotation}\noindent\textsc{Key:} #1\\
  \textsc{Annotation:}\ \input{#1}\end{quotation}}{}}
\providecommand{\bibAnnote}[2]{%
  \begin{quotation}\noindent\textsc{Key:} #1\\
  \textsc{Annotation:}\ #2\end{quotation}}

\bibitem{LDA}
J.~P. Perdew and A.~Zunger, \emph{Self-interaction correction to
  density-functional approximations for many-electron systems}, Phys.\ Rev.\ B
  \textbf{23}, 5048--5079 (1981), \doi{10.1103/PhysRevB.23.5048}.
\input{LDA}

\bibitem{PBE}
J.~P. Perdew, K.~Burke, and M.~Ernzerhof, \emph{Generalized Gradient
  Approximation Made Simple}, Phys.\ Rev.\ Lett. \textbf{77}, 3865--3868
  (1996), \doi{10.1103/PhysRevLett.77.3865}.
\input{PBE}

\bibitem{Perdew_SCAN_PRL_2015}
J.~Sun, A.~Ruzsinszky, and J.~P. Perdew, \emph{Strongly Constrained and
  Appropriately Normed Semilocal Density Functional}, Phys.\ Rev.\ Lett.
  \textbf{115}, 036402 (2015), \doi{10.1103/PhysRevLett.115.036402}.
\input{Perdew_SCAN_PRL_2015}

\bibitem{Oses_CMS_2023}
C.~Oses, M.~Esters, D.~Hicks, S.~Divilov, H.~Eckert, R.~Friedrich, M.~J. Mehl,
  A.~Smolyanyuk, X.~Campilongo, A.~{van de Walle}, J.~Schroers, A.~G. Kusne,
  I.~Takeuchi, E.~Zurek, M.~{Buongiorno Nardelli}, M.~Fornari, Y.~Lederer,
  O.~Levy, C.~Toher, and S.~Curtarolo, \emph{{aflow++}: A {C++} framework for
  autonomous materials design}, Comput.\ Mater.\ Sci. \textbf{217}, 111889
  (2023), \doi{10.1016/j.commatsci.2022.111889}.
\input{Oses_CMS_2023}

\bibitem{Esters_CMS_2023}
M.~Esters, C.~Oses, S.~Divilov, H.~Eckert, R.~Friedrich, D.~Hicks, M.~J. Mehl,
  F.~Rose, A.~Smolyanyuk, A.~Calzolari, X.~Campilongo, C.~Toher, and
  S.~Curtarolo, \emph{{aflow.org}: A web ecosystem of databases, software and
  tools}, Comput.\ Mater.\ Sci. \textbf{216}, 111808 (2023),
  \doi{10.1016/j.commatsci.2022.111808}.
\input{Esters_CMS_2023}

\bibitem{curtarolo:art75}
S.~Curtarolo, W.~Setyawan, S.~Wang, J.~Xue, K.~Yang, R.~H. Taylor, L.~J.
  Nelson, G.~L.~W. Hart, S.~Sanvito, M.~{Buongiorno Nardelli}, N.~Mingo, and
  O.~Levy, \emph{{AFLOWLIB.ORG}: A distributed materials properties repository
  from high-throughput {\it ab initio} calculations}, Comput.\ Mater.\ Sci.
  \textbf{58}, 227--235 (2012), \doi{10.1016/j.commatsci.2012.02.002}.
\input{curtarolo:art75}

\bibitem{materialsproject.org}
A.~Jain, G.~Hautier, C.~J. Moore, S.~P. Ong, C.~C. Fischer, T.~Mueller, K.~A.
  Persson, and G.~Ceder, \emph{A high-throughput infrastructure for density
  functional theory calculations}, Comput.\ Mater.\ Sci. \textbf{50},
  2295--2310 (2011), \doi{10.1016/j.commatsci.2011.02.023}.
\input{materialsproject.org}

\bibitem{oqmd.org}
J.~E. Saal, S.~Kirklin, M.~Aykol, B.~Meredig, and C.~Wolverton, \emph{Materials
  Design and Discovery with High-Throughput Density Functional Theory: The
  {O}pen {Q}uantum {M}aterials {D}atabase ({OQMD})}, JOM \textbf{65},
  1501--1509 (2013), \doi{10.1007/s11837-013-0755-4}.
\input{oqmd.org}

\bibitem{Kirklin_NPJCM_2015}
S.~Kirklin, J.~E. Saal, B.~Meredig, A.~Thompson, J.~W. Doak, M.~Aykol,
  S.~R{\"{u}}hl, and C.~Wolverton, \emph{The {Open} {Quantum} {Materials}
  {Database} ({OQMD}): assessing the accuracy of {DFT} formation energies},
  npj\ Comput.\ Mater. \textbf{1}, 15010 (2015),
  \doi{10.1038/npjcompumats.2015.10}.
\input{Kirklin_NPJCM_2015}

\bibitem{nomadMRS}
C.~Draxl and M.~Scheffler, \emph{{NOMAD}: The {FAIR} concept for big
  data-driven materials science}, MRS\ Bull. \textbf{43}, 676--682 (2018),
  \doi{10.1557/mrs.2018.208}.
\input{nomadMRS}

\bibitem{ase}
S.~R. Bahn and K.~W. Jacobsen, \emph{An object-oriented scripting interface to
  a legacy electronic structure code}, Comput.\ Sci.\ Eng. \textbf{4}, 56--66
  (2002), \doi{10.1109/5992.998641}.
\input{ase}

\bibitem{cmr_repository}
D.~D. Landis, J.~S. Hummelsh{\o}j, S.~Nestorov, J.~Greeley, M.~Du{\l}ak,
  T.~Bligaard, J.~K. N{\o}rskov, and K.~W. Jacobsen, \emph{The Computational
  Materials Repository}, Comput.\ Sci.\ Eng. \textbf{14}, 51--57 (2012),
  \doi{10.1109/MCSE.2012.16}.
\input{cmr_repository}

\bibitem{Pizzi_AiiDA_2016}
G.~Pizzi, A.~Cepellotti, R.~Sabatini, N.~Marzari, and B.~Kozinsky,
  \emph{{AiiDA}: automated interactive infrastructure and database for
  computational science}, Comput.\ Mater.\ Sci. \textbf{111}, 218--230 (2016).
\input{Pizzi_AiiDA_2016}

\bibitem{Dirac1929}
P.~A.~M. Dirac, \emph{Quantum Mechanics of Many-Electron Systems}, Proc.\ R.\
  Soc.\ A\ Math.\ Phys.\ Eng.\ Sci. \textbf{123}, 714--733 (1929),
  \doi{10.1098/rspa.1929.0094}.
\input{Dirac1929}

\bibitem{curtarolo:art176}
M.~J. Mehl, M.~Ronquillo, D.~Hicks, M.~Esters, C.~Oses, R.~Friedrich,
  A.~Smolyanyuk, E.~Gossett, D.~Finkenstadt, and S.~Curtarolo, \emph{{The Tin
  Pest Problem as a Test of Density Functionals Using High-Throughput
  Calculations}}, Phys.\ Rev.\ Materials \textbf{5}, 083608 (2021),
  \doi{10.1103/PhysRevMaterials.5.083608}.
\input{curtarolo:art176}

\bibitem{SarkerHarrington_HEC_NCOMMS_2018_etal}
P.~Sarker, T.~Harrington, et~al., \emph{High-entropy high-hardness metal
  carbides discovered by entropy descriptors}, Nat.\ Commun. \textbf{9}, 4980
  (2018), \doi{10.1038/s41467-018-07160-7}.
\input{SarkerHarrington_HEC_NCOMMS_2018_etal}

\bibitem{Oses_NatureReview_2020}
C.~Oses, C.~Toher, and S.~Curtarolo, \emph{High-entropy ceramics}, Nat.\ Rev.\
  Mater. \textbf{5}, 295--309 (2020), \doi{10.1038/s41578-019-0170-8}.
\input{Oses_NatureReview_2020}

\bibitem{Gil_PCCP_2022}
J.~Gil and T.~Oda, \emph{Correction methods for first-principles calculations
  of the solution enthalpy of gases and compounds in liquid metals}, Phys.\
  Chem.\ Chem.\ Phys. \textbf{24}, 757--770 (2022), \doi{10.1039/D1CP02450G}.
\input{Gil_PCCP_2022}

\bibitem{Puthirath_Balan_NNANO_2018}
A.~Puthirath~Balan, S.~Radhakrishnan, C.~F. Woellner, S.~K. Sinha, L.~Deng,
  C.~d.~l. Reyes, B.~M. Rao, M.~Paulose, R.~Neupane, A.~Apte, V.~Kochat,
  R.~Vajtai, A.~R. Harutyunyan, C.-W. Chu, G.~Costin, D.~S. Galvao, A.~A.
  Mart\'{i}, P.~A. van Aken, O.~K. Varghese, C.~S. Tiwary, A.~Malie Madom
  Ramaswamy~Iyer, and P.~M. Ajayan, \emph{Exfoliation of a non-van der {Waals}
  material from iron ore hematite}, Nat.\ Nanotechnol. \textbf{13}, 602--609
  (2018), \doi{10.1038/s41565-018-0134-y}.
\input{Puthirath_Balan_NNANO_2018}

\bibitem{Friedrich_NanoLett_2022}
R.~Friedrich, M.~Ghorbani-Asl, S.~Curtarolo, and A.~V. Krasheninnikov,
  \emph{Data-{Driven} {Quest} for {Two}-{Dimensional} {Non}-van der {Waals}
  {Materials}}, Nano\ Lett. \textbf{22}, 989--997 (2022),
  \doi{10.1021/acs.nanolett.1c03841}.
\input{Friedrich_NanoLett_2022}

\bibitem{Barnowsky_AdvElMats_2023}
T.~Barnowsky, A.~V. Krasheninnikov, and R.~Friedrich, \emph{A {New} {Group} of
  {2D} {Non}-van der {Waals} {Materials} with {Ultra} {Low} {Exfoliation}
  {Energies}}, Adv.\ Electron.\ Mater. \textbf{9}, 2201112 (2023),
  \doi{10.1002/aelm.202201112}.
\input{Barnowsky_AdvElMats_2023}

\bibitem{Balan_MatTod_2022}
A.~P. Balan, A.~B. Puthirath, S.~Roy, G.~Costin, E.~F. Oliveira, M.~A. S.~R.
  Saadi, V.~Sreepal, R.~Friedrich, P.~Serles, A.~Biswas, S.~A. Iyengar,
  N.~Chakingal, S.~Bhattacharyya, S.~K. Saju, S.~C. Pardo, L.~M. Sassi,
  T.~Filleter, A.~Krasheninnikov, D.~S. Galvao, R.~Vajtai, R.~R. Nair, and
  P.~M. Ajayan, \emph{Non-van der {Waals} quasi-{2D} materials; recent advances
  in synthesis, emergent properties and applications}, Mater.\ Today
  \textbf{58}, 164 (2022), \doi{10.1016/j.mattod.2022.07.007}.
\input{Balan_MatTod_2022}

\bibitem{Kaur_AdvMat_2022}
H.~Kaur and J.~N. Coleman, \emph{Liquid-{Phase} {Exfoliation} of {Nonlayered}
  {Non}-{Van}-{Der}-{Waals} {Crystals} into {Nanoplatelets}}, Adv.\ Mater.
  \textbf{34}, 2202164 (2022), \doi{10.1002/adma.202202164}.
\input{Kaur_AdvMat_2022}

\bibitem{Wang_Ceder_GGAU_PRB_2006}
L.~Wang, T.~Maxisch, and G.~Ceder, \emph{Oxidation energies of transition metal
  oxides within the GGA+U framework}, Phys.\ Rev.\ B \textbf{73}, 195107
  (2006), \doi{10.1103/PhysRevB.73.195107}.
\input{Wang_Ceder_GGAU_PRB_2006}

\bibitem{Lany_FERE_2008}
S.~Lany, \emph{Semiconductor thermochemistry in density functional
  calculations}, Phys.\ Rev.\ B \textbf{78}, 245207 (2008),
  \doi{10.1103/PhysRevB.78.245207}.
\input{Lany_FERE_2008}

\bibitem{Jain_GGAU_PRB_2011}
A.~Jain, G.~Hautier, S.~P. Ong, C.~J. Moore, C.~C. Fischer, K.~A. Persson, and
  G.~Ceder, \emph{Formation enthalpies by mixing GGA and GGA+U calculations},
  Phys.\ Rev.\ B \textbf{84}, 045115 (2011), \doi{10.1103/PhysRevB.84.045115}.
\input{Jain_GGAU_PRB_2011}

\bibitem{Stevanovic_FERE_2012}
V.~Stevanovi{\'{c}}, S.~Lany, X.~Zhang, and A.~Zunger, \emph{Correcting density
  functional theory for accurate predictions of compound enthalpies of
  formation: Fitted elemental-phase reference energies}, Phys.\ Rev.\ B
  \textbf{85}, 115104 (2012), \doi{10.1103/PhysRevB.85.115104}.
\input{Stevanovic_FERE_2012}

\bibitem{Yan_formation_PRB_2013}
J.~Yan, J.~S. Hummelsh{\o}j, and J.~K. N{\o}rskov, \emph{Formation energies of
  group {I} and {II} metal oxides using random phase approximation}, Phys.\
  Rev.\ B \textbf{87}, 075207 (2013), \doi{10.1103/PhysRevB.87.075207}.
\input{Yan_formation_PRB_2013}

\bibitem{Jauho_PRB_2015}
T.~S. Jauho, T.~Olsen, T.~Bligaard, and K.~S. Thygesen, \emph{Improved
  description of metal oxide stability: {Beyond} the random phase approximation
  with renormalized kernels}, Phys.\ Rev.\ B \textbf{92}, 115140 (2015),
  \doi{10.1103/PhysRevB.92.115140}.
\input{Jauho_PRB_2015}

\bibitem{Zhang_NPJCM_2018}
Y.~Zhang, D.~A. Kitchaev, J.~Yang, T.~Chen, S.~T. Dacek, R.~A.
  Sarmiento-P{\'{e}}rez, M.~A.~L. Marques, H.~Peng, G.~Ceder, J.~P. Perdew, and
  J.~Sun, \emph{Efficient first-principles prediction of solid stability:
  {Towards} chemical accuracy}, npj\ Comput.\ Mater. \textbf{4}, 9 (2018),
  \doi{10.1038/s41524-018-0065-z}.
\input{Zhang_NPJCM_2018}

\bibitem{Isaacs_PRM_2018}
E.~B. Isaacs and C.~Wolverton, \emph{Performance of the strongly constrained
  and appropriately normed density functional for solid-state materials},
  Phys.\ Rev.\ Materials \textbf{2}, 063801 (2018),
  \doi{10.1103/PhysRevMaterials.2.063801}.
\input{Isaacs_PRM_2018}

\bibitem{Friedrich_CCE_2019}
R.~Friedrich, D.~Usanmaz, C.~Oses, A.~Supka, M.~Fornari, M.~{Buongiorno
  Nardelli}, C.~Toher, and S.~Curtarolo, \emph{Coordination corrected ab initio
  formation enthalpies}, npj\ Comput.\ Mater. \textbf{5}, 59 (2019),
  \doi{10.1038/s41524-019-0192-1}.
\input{Friedrich_CCE_2019}

\bibitem{Kubaschewski_MTC_1993}
O.~Kubaschewski, C.~B. Alcock, and P.~J. Spencer, \emph{Materials
  Thermochemistry} (Pergamon Press, Oxford, UK, 1993), 6th edn.
\input{Kubaschewski_MTC_1993}

\bibitem{Chase_NIST_JANAF_thermochem_tables_1998}
M.~W. {Chase,~Jr.}, \emph{{NIST}-{JANAF} Thermochemical Tables} (American
  Chemical Society and American Institute of Physics for the National Institute
  of Standards and Technology, Woodbury, NY, 1998), 4th edn.
\input{Chase_NIST_JANAF_thermochem_tables_1998}

\bibitem{Barin_1995}
I.~Barin, \emph{Thermochemical Data of Pure Substances} (VCH, Weinheim, 1995),
  3rd edn.
\input{Barin_1995}

\bibitem{Wagman_NBS_thermodyn_tables_1982}
D.~D. Wagman, W.~H. Evans, V.~B. Parker, R.~H. Schumm, I.~Halow, S.~M. Bailey,
  K.~L. Churney, and R.~L. Nuttall, \emph{The {NBS} tables of chemical
  thermodynamic properties}, J. Phys. Chem. Ref. Data \textbf{11}, Supplement
  No. 2 (1982).
\input{Wagman_NBS_thermodyn_tables_1982}

\bibitem{Pederson_JCP_2014}
M.~R. Pederson, A.~Ruzsinszky, and J.~P. Perdew, \emph{Communication:
  {Self}-interaction correction with unitary invariance in density functional
  theory}, J.\ Chem.\ Phys. \textbf{140}, 121103 (2014),
  \doi{10.1063/1.4869581}.
\input{Pederson_JCP_2014}

\bibitem{Yang_PRA_2017}
Z.-h. Yang, M.~R. Pederson, and J.~P. Perdew, \emph{Full self-consistency in
  the {Fermi}-orbital self-interaction correction}, Phys.\ Rev.\ A \textbf{95},
  052505 (2017), \doi{10.1103/PhysRevA.95.052505}.
\input{Yang_PRA_2017}

\bibitem{Kao_JCP_2017}
D.-y. Kao, K.~Withanage, T.~Hahn, J.~Batool, J.~Kortus, and K.~Jackson,
  \emph{Self-consistent self-interaction corrected density functional theory
  calculations for atoms using {Fermi}-{L{\"{o}}wdin} orbitals: {Optimized}
  {Fermi}-orbital descriptors for {Li}-{Kr}}, J.\ Chem.\ Phys. \textbf{147},
  164107 (2017), \doi{10.1063/1.4996498}.
\input{Kao_JCP_2017}

\bibitem{Schwalbe_JCC_Fermi-Lowdin_2018}
S.~Schwalbe, T.~Hahn, S.~Liebing, K.~Trepte, and J.~Kortus,
  \emph{Fermi-{L{\"{o}}wdin} Orbital Self-interaction Corrected Density
  Functional Theory: {Ionization} Potentials and Enthalpies of Formation}, J.\
  Comput.\ Chem. \textbf{39}, 2463--2471 (2018), \doi{10.1002/jcc.25586}.
\input{Schwalbe_JCC_Fermi-Lowdin_2018}

\bibitem{Pozzo_PRB_2008}
M.~Pozzo and D.~Alf{\'{e}}, \emph{Structural properties and enthalpy of
  formation of magnesium hydride from quantum {Monte} {Carlo} calculations},
  Phys.\ Rev.\ B \textbf{77}, 104103 (2008), \doi{10.1103/PhysRevB.77.104103}.
\input{Pozzo_PRB_2008}

\bibitem{Mao_QMC_2011}
G.~Mao, X.~Hu, X.~Wu, Y.~Dai, S.~Chu, and J.~Deng, \emph{Benchmark {Quantum}
  {Monte} {Carlo} calculation of the enthalpy of formation of {Mg}{H}$_2$},
  Int.\ J.\ of\ Hydrogen\ Energy \textbf{36}, 8388--8391 (2011),
  \doi{10.1016/j.ijhydene.2011.04.093}.
\input{Mao_QMC_2011}

\bibitem{Isaacs_PRB_2022}
E.~B. Isaacs, H.~Shin, A.~Annaberdiyev, C.~Wolverton, L.~Mitas, A.~Benali, and
  O.~Heinonen, \emph{Assessing the accuracy of compound formation energies with
  quantum {Monte} {Carlo}}, Phys.\ Rev.\ B \textbf{105}, 224110 (2022),
  \doi{10.1103/PhysRevB.105.224110}.
\input{Isaacs_PRB_2022}

\bibitem{Wang_SciRep_2021}
A.~Wang, R.~Kingsbury, M.~McDermott, M.~Horton, A.~Jain, S.~P. Ong,
  S.~Dwaraknath, and K.~A. Persson, \emph{A framework for quantifying
  uncertainty in {DFT} energy corrections}, Sci.\ Rep. \textbf{11}, 15496
  (2021), \doi{10.1038/s41598-021-94550-5}.
\input{Wang_SciRep_2021}

\bibitem{Wolverton_DFTUenthalpies_prb_2014}
M.~Aykol and C.~Wolverton, \emph{Local environment dependent GGA+U method for
  accurate thermochemistry of transition metal compounds}, Phys.\ Rev.\ B
  \textbf{90}, 115105 (2014), \doi{10.1103/PhysRevB.90.115105}.
\input{Wolverton_DFTUenthalpies_prb_2014}

\bibitem{Artrith_PRM_2022}
N.~Artrith, J.~A. Garrido~Torres, A.~Urban, and M.~S. Hybertsen,
  \emph{Data-driven approach to parameterize SCAN$+{U}$ for an accurate
  description of $3d$ transition metal oxide thermochemistry}, Phys.\ Rev.\
  Materials \textbf{6}, 035003 (2022), \doi{10.1103/PhysRevMaterials.6.035003}.
\input{Artrith_PRM_2022}

\bibitem{Kingsbury_NPJCM_2022}
R.~S. Kingsbury, A.~S. Rosen, A.~S. Gupta, J.~M. Munro, S.~P. Ong, A.~Jain,
  S.~Dwaraknath, M.~K. Horton, and K.~A. Persson, \emph{A flexible and scalable
  scheme for mixing computed formation energies from different levels of
  theory}, npj\ Comput.\ Mater. \textbf{8}, 195 (2022),
  \doi{10.1038/s41524-022-00881-w}.
\input{Kingsbury_NPJCM_2022}

\bibitem{Gong_JACSAu_2022}
S.~Gong, S.~Wang, T.~Xie, W.~H. Chae, R.~Liu, Y.~Shao-Horn, and J.~C. Grossman,
  \emph{Calibrating {DFT} {Formation} {Enthalpy} {Calculations} by
  {Multifidelity} {Machine} {Learning}}, JACS Au \textbf{2}, 1964--1977 (2022),
  \doi{10.1021/jacsau.2c00235}.
\input{Gong_JACSAu_2022}

\bibitem{Friedrich_PRM_2021}
R.~Friedrich, M.~Esters, C.~Oses, S.~Ki, M.~J. Brenner, D.~Hicks, M.~J. Mehl,
  C.~Toher, and S.~Curtarolo, \emph{Automated coordination corrected enthalpies
  with {AFLOW}-{CCE}}, Phys.\ Rev.\ Materials \textbf{5}, 043803 (2021),
  \doi{10.1103/PhysRevMaterials.5.043803}.
\input{Friedrich_PRM_2021}

\bibitem{Blanco_jmolstrthch_1996}
M.~A. Blanco, A.~M. Pend{\'{a}}s, E.~Francisco, J.~M. Recio, and R.~Franco,
  \emph{Thermodynamical properties of solids from microscopic theory:
  Applications to MgF$_2$ and Al$_2$O$_3$}, J.\ Mol.\ Struct.:\ Theochem
  \textbf{368}, 245--255 (1996), \doi{10.1016/S0166-1280(96)90571-0}.
\input{Blanco_jmolstrthch_1996}

\bibitem{BlancoGIBBS2004}
M.~A. Blanco, E.~Francisco, and V.~Lua{\~{n}}a, \emph{{GIBBS}:
  isothermal-isobaric thermodynamics of solids from energy curves using a
  quasi-harmonic Debye model}, Comput.\ Phys.\ Commun. \textbf{158}, 57--72
  (2004), \doi{10.1016/j.comphy.2003.12.001}.
\input{BlancoGIBBS2004}

\bibitem{curtarolo:art96}
C.~Toher, J.~J. Plata, O.~Levy, M.~{de~Jong}, M.~Asta, M.~{Buongiorno
  Nardelli}, and S.~Curtarolo, \emph{High-throughput computational screening of
  thermal conductivity, {D}ebye temperature, and {G}r{\"{u}}neisen parameter
  using a quasiharmonic {D}ebye model}, Phys.\ Rev.\ B \textbf{90}, 174107
  (2014), \doi{10.1103/PhysRevB.90.174107}.
\input{curtarolo:art96}

\bibitem{curtarolo:art115}
C.~Toher, C.~Oses, J.~J. Plata, D.~Hicks, F.~Rose, O.~Levy, M.~{de Jong},
  M.~Asta, M.~Fornari, M.~{Buongiorno Nardelli}, and S.~Curtarolo,
  \emph{Combining the {AFLOW} {GIBBS} and elastic libraries to efficiently and
  robustly screen thermomechanical properties of solids}, Phys.\ Rev.\
  Materials \textbf{1}, 015401 (2017), \doi{10.1103/PhysRevMaterials.1.015401}.
\input{curtarolo:art115}

\bibitem{Poirier_Earth_Interior_2000}
J.-P. Poirier, \emph{Introduction to the Physics of the Earth's Interior}
  (Cambridge University Press, 2000), 2nd edn.
\input{Poirier_Earth_Interior_2000}

\bibitem{curtarolo:art125}
J.~J. Plata, P.~Nath, D.~Usanmaz, J.~Carrete, C.~Toher, M.~{de Jong}, M.~D.
  Asta, M.~Fornari, M.~{Buongiorno Nardelli}, and S.~Curtarolo, \emph{An
  efficient and accurate framework for calculating lattice thermal conductivity
  of solids: {AFLOW}-{AAPL} {Au}tomatic {A}nharmonic {P}honon {Li}brary}, npj\
  Comput.\ Mater. \textbf{3}, 45 (2017), \doi{10.1038/s41524-017-0046-7}.
\input{curtarolo:art125}

\bibitem{Sun_NChem_2016}
J.~Sun, R.~C. Remsing, Y.~Zhang, Z.~Sun, A.~Ruzsinszky, H.~Peng, Z.~Yang,
  A.~Paul, U.~Waghmare, X.~Wu, M.~L. Klein, and J.~P. Perdew, \emph{Accurate
  first-principles structures and energies of diversely bonded systems from an
  efficient density functional}, Nat.\ Chem. \textbf{8}, 9 (2016),
  \doi{10.1038/nchem.2535}.
\input{Sun_NChem_2016}

\bibitem{curtarolo:art104}
C.~E. Calderon, J.~J. Plata, C.~Toher, C.~Oses, O.~Levy, M.~Fornari, A.~Natan,
  M.~J. Mehl, G.~L.~W. Hart, M.~{Buongiorno Nardelli}, and S.~Curtarolo,
  \emph{The {AFLOW} standard for high-throughput materials science
  calculations}, Comput.\ Mater.\ Sci. \textbf{108}, 233--238 (2015),
  \doi{10.1016/j.commatsci.2015.07.019}.
\input{curtarolo:art104}

\bibitem{DFT}
W.~Kohn and L.~J. Sham, \emph{Self-Consistent Equations Including Exchange and
  Correlation Effects}, Phys.\ Rev. \textbf{140}, A1133 (1965),
  \doi{10.1103/PhysRev.140.A1133}.
\input{DFT}

\bibitem{von_Barth_JPCSS_LSDA_1972}
U.~von Barth and L.~Hedin, \emph{A local exchange-correlation potential for the
  spin polarized case: I}, J.\ Phys.\ C:\ Solid\ State\ Phys. \textbf{5}, 1629
  (1972), \doi{10.1088/0022-3719/5/13/012}.
\input{von_Barth_JPCSS_LSDA_1972}

\bibitem{curtarolo:art53}
O.~Levy, R.~V. Chepulskii, G.~L.~W. Hart, and S.~Curtarolo, \emph{The New Face
  of {Rh}odium Alloys: Revealing Ordered Structures from First Principles}, J.\
  Am.\ Chem.\ Soc. \textbf{132}, 833--837 (2010), \doi{10.1021/ja908879y}.
\input{curtarolo:art53}

\bibitem{curtarolo:art57}
O.~Levy, G.~L.~W. Hart, and S.~Curtarolo, \emph{Structure maps for {h}cp metals
  from first-principles calculations}, Phys.\ Rev.\ B \textbf{81}, 174106
  (2010), \doi{10.1103/PhysRevB.81.174106}.
\input{curtarolo:art57}

\bibitem{curtarolo:art63}
O.~Levy, M.~Jahn{\'{a}}tek, R.~V. Chepulskii, G.~L.~W. Hart, and S.~Curtarolo,
  \emph{Ordered Structures in {Rh}enium Binary Alloys from First-Principles
  Calculations}, J.\ Am.\ Chem.\ Soc. \textbf{133}, 158--163 (2011),
  \doi{10.1021/ja1091672}.
\input{curtarolo:art63}

\bibitem{aflowPAPER}
S.~Curtarolo, W.~Setyawan, G.~L.~W. Hart, M.~Jahn{\'{a}}tek, R.~V. Chepulskii,
  R.~H. Taylor, S.~Wang, J.~Xue, K.~Yang, O.~Levy, M.~J. Mehl, H.~T. Stokes,
  D.~O. Demchenko, and D.~Morgan, \emph{{AFLOW}: An automatic framework for
  high-throughput materials discovery}, Comput.\ Mater.\ Sci. \textbf{58},
  218--226 (2012), \doi{10.1016/j.commatsci.2012.02.005}.
\input{aflowPAPER}

\bibitem{curtarolo:art110}
K.~Yang, C.~Oses, and S.~Curtarolo, \emph{Modeling Off-Stoichiometry Materials
  with a High-Throughput \textit{Ab-Initio} Approach}, Chem.\ Mater.
  \textbf{28}, 6484--6492 (2016), \doi{10.1021/acs.chemmater.6b01449}.
\input{curtarolo:art110}

\bibitem{aflowPI}
A.~R. Supka, T.~E. Lyons, L.~S.~I. Liyanage, P.~{D'{A}mico},
  R.~{Al~Rahal~Al~Orabi}, S.~Mahatara, P.~Gopal, C.~Toher, D.~Ceresoli,
  A.~Calzolari, S.~Curtarolo, M.~{Buongiorno Nardelli}, and M.~Fornari,
  \emph{{\small AFLOW}$\pi$: A minimalist approach to high-throughput
  \textit{ab initio} calculations including the generation of tight-binding
  hamiltonians}, Comput.\ Mater.\ Sci. \textbf{136}, 76--84 (2017),
  \doi{10.1016/j.commatsci.2017.03.055}.
\input{aflowPI}

\bibitem{vasp}
G.~Kresse and J.~Furthm{\"{u}}ller, \emph{Efficient iterative schemes for {\it
  ab initio} total-energy calculations using a plane-wave basis set}, Phys.\
  Rev.\ B \textbf{54}, 11169--11186 (1996), \doi{10.1103/PhysRevB.54.11169}.
\input{vasp}

\bibitem{Sun_CoM_2017}
W.~Sun, A.~Holder, B.~Orva{\~{n}}anos, E.~Arca, A.~Zakutayev, S.~Lany, and
  G.~Ceder, \emph{Thermodynamic {Routes} to {Novel} {Metastable}
  {Nitrogen}-{Rich} {Nitrides}}, Chem.\ Mater. \textbf{29}, 6936--6946 (2017),
  \doi{10.1021/acs.chemmater.7b02399}.
\input{Sun_CoM_2017}

\bibitem{Sun_NMAT_2019}
W.~Sun, C.~J. Bartel, E.~Arca, S.~R. Bauers, B.~Matthews, B.~Orva{\~n}anos,
  B.-R. Chen, M.~F. Toney, L.~T. Schelhas, W.~Tumas, J.~Tate, A.~Zakutayev,
  S.~Lany, A.~M. Holder, and G.~Ceder, \emph{A map of the inorganic ternary
  metal nitrides}, Nat.\ Mater. \textbf{18}, 732--739 (2019),
  \doi{10.1038/s41563-019-0396-2}.
\input{Sun_NMAT_2019}

\bibitem{curtarolo:art140}
P.~Sarker, T.~Harrington, C.~Toher, C.~Oses, M.~Samiee, J.-P. Maria, D.~W.
  Brenner, K.~S. Vecchio, and S.~Curtarolo, \emph{High-entropy high-hardness
  metal carbides discovered by entropy descriptors}, Nat.\ Commun. \textbf{9},
  4980 (2018), \doi{10.1038/s41467-018-07160-7}.
\input{curtarolo:art140}

\bibitem{Ranade_JPCB_2000}
M.~R. Ranade, F.~Tessier, A.~Navrotsky, V.~J. Leppert, S.~H. Risbud, F.~J.
  DiSalvo, and C.~M. Balkas, \emph{Enthalpy of {Formation} of {Gallium}
  {Nitride}}, J.\ Phys.\ Chem.\ B \textbf{104}, 4060--4063 (2000),
  \doi{10.1021/jp993752s}.
\input{Ranade_JPCB_2000}

\bibitem{McHale_ChemMater_1997}
J.~M. McHale, A.~Navrotsky, G.~R. Kowach, V.~E. Balbarin, and F.~J. DiSalvo,
  \emph{Energetics of {Ternary} {Nitrides}: Li-Ca-Zn-N and Ca-Ta-N Systems},
  Chem.\ Mater. \textbf{9}, 1538--1546 (1997), \doi{10.1021/cm970244r}.
\input{McHale_ChemMater_1997}

\bibitem{McHale_ChemMater_1999}
J.~M. McHale, A.~Navrotsky, and F.~J. DiSalvo, \emph{Energetics of {Ternary}
  {Nitride} {Formation} in the (Li,Ca)-(B,Al)-N {System}}, Chem.\ Mater.
  \textbf{11}, 1148--1152 (1999), \doi{10.1021/cm981096n}.
\input{McHale_ChemMater_1999}

\bibitem{curtarolo:art121}
M.~J. Mehl, D.~Hicks, C.~Toher, O.~Levy, R.~M. Hanson, G.~L.~W. Hart, and
  S.~Curtarolo, \emph{The {AFLOW} Library of Crystallographic Prototypes: Part
  1}, Comput.\ Mater.\ Sci. \textbf{136}, S1--S828 (2017),
  \doi{10.1016/j.commatsci.2017.01.017}.
\input{curtarolo:art121}

\bibitem{curtarolo:art145}
D.~Hicks, M.~J. Mehl, E.~Gossett, C.~Toher, O.~Levy, R.~M. Hanson, G.~L.~W.
  Hart, and S.~Curtarolo, \emph{The {AFLOW} Library of Crystallographic
  Prototypes: Part 2}, Comput.\ Mater.\ Sci. \textbf{161}, S1--S1011 (2019),
  \doi{10.1016/j.commatsci.2018.10.043}.
\input{curtarolo:art145}

\bibitem{curtarolo:art173}
D.~Hicks, M.~J. Mehl, M.~Esters, C.~Oses, O.~Levy, G.~L.~W. Hart, C.~Toher, and
  S.~Curtarolo, \emph{The {AFLOW} Library of Crystallographic Prototypes: Part
  3}, Comput.\ Mater.\ Sci. \textbf{199}, 110450 (2021),
  \doi{10.1016/j.commatsci.2021.110450}.
\input{curtarolo:art173}

\bibitem{curtarolo:art135}
D.~Hicks, C.~Oses, E.~Gossett, G.~Gomez, R.~H. Taylor, C.~Toher, M.~J. Mehl,
  O.~Levy, and S.~Curtarolo, \emph{\textit{AFLOW-SYM}: platform for the
  complete, automatic and self-consistent symmetry analysis of crystals}, Acta\
  Crystallogr.\ Sect.\ A \textbf{74}, 184--203 (2018),
  \doi{10.1107/S2053273318003066}.
\input{curtarolo:art135}

\end{thebibliography}
\end{document}